\documentclass[acmsmall,screen]{acmart}

\AtBeginDocument{%
  }

\usepackage{amsmath,amsfonts}
\usepackage{algorithm} 
\usepackage{algpseudocode} 
\usepackage{graphicx}
\usepackage{subcaption}
\usepackage{textcomp}
\usepackage{xcolor}
\usepackage{xspace}
\usepackage{enumitem}
\usepackage{balance}

\newcommand{\ourmethod}{\texttt{BODE-GEN}\xspace}

\begin{document}

\title{An Exploratory Study of Bayesian Prompt Optimization for\\Test-Driven Code Generation with Large Language Models
}

\author{Shlok Tomar}
\affiliation{%
  \institution{Washington State University}
  \country{USA}
}
\email{shlok.tomar@wsu.edu}
\author{Aryan Deshwal}
\affiliation{%
  \institution{University of Minnesota}
  \country{USA}
}
\email{adeshwal@umn.edu}
\author{Ethan Villalovoz}
\affiliation{%
  \institution{Washington State University}
  \country{USA}
}
\email{ethan.villalovoz@wsu.edu}
\author{Mattia Fazzini}
\affiliation{%
  \institution{University of Minnesota}
  \country{USA}
}
\email{mfazzini@umn.edu}
\author{Haipeng Cai}
\affiliation{%
  \institution{University at Buffalo}
  \country{USA}
}
\email{haipengc@buffalo.edu}
\author{Janardhan Rao Doppa}
\affiliation{%
  \institution{Washington State University}
  \country{USA}
}
\email{jana.doppa@wsu.edu}







\begin{abstract}
We consider the task of generating functionally correct code using large language models (LLMs). The correctness of generated code is influenced by the prompt used to query the given base LLM. We formulate the problem of finding the appropriate prompt as combinatorial search process and propose a Bayesian optimization (BO) approach referred to as {\em BO for Code GENeration (BODE-GEN)}. BODE-GEN performs an adaptive data-driven search over prompts guided by training data in the form of prompts tried and the functional accuracy of the generated code over a set of given test cases. The key insight is to perform BO in continuous embedding space by using an auxiliary LLM to bridge the gap between discrete prompt space and continuous embedding space. We leverage two synergistic ideas, namely, random projections and dimensionality scaled priors, to build effective Gaussian process based surrogate models over the high-dimensional embedding space. Our experiments on the HumanEval+ benchmark using multiple base LLMs show that BODE-GEN can improve performance in terms of code generation accuracy compared to fixed prompts and manual prompt engineering. Additionally, we demonstrate that BODE-GEN is sample-efficient, requiring relatively few iterations of BO to demonstrate improvements in code accuracy. \end{abstract}



\maketitle



\section{Introduction}
Large language models (LLMs) have emerged as transformative tools in various domains, including software development. Their ability to assist with code-related tasks has positioned them as indispensable coding assistants for software developers today\cite{nam2024using,coignion2024performance,chen2021evaluating}. 
However, as developers increasingly rely on code generated by LLMs, the functional correctness of this code has become an important factor in ensuring the overall quality of software products. 

In modern software development, the software supply chain comprises of various components, including low-level systems software, application frameworks, third-party libraries, and build tools. These components are often inter-dependent and any flaw in one can propagate through the entire system, leading to significant and widespread challenges~\cite{enck2022top,ellison2010evaluating}. When developers incorporate LLM-generated code into these components, the functional correctness of the code becomes vital. Incorrect code generated by LLMs can introduce subtle bugs that are difficult to detect and diagnose, potentially causing cascading side-effects for the entire software supply chain~\cite{balayn2024empirical}. Therefore, ensuring the correctness of LLM-generated code before its incorporation into the software development process is of paramount importance.

Recent studies have highlighted that LLMs can produce incorrect code, for various reasons,  including hallucinations and insufficient understanding of coding tasks \cite{tian2024codehalu,liu2024exploring,nam2024using}. To address this challenge, some potential solutions include better pre-training datasets and improved training/fine-tuning methods to create high-performing LLMs specifically for coding tasks \cite{guo2024deepseek}. There is also an inherent trade-off between resource cost and performance of LLMs. On one end of the spectrum, large LLMs require huge amount of high-quality training data and compute resources for both training and inference \cite{wang2021codex,gao2021making}. On the other end, training and inference with smaller LLMs is feasible and computationally cheap, but these models may struggle to generalize well to handle the complexities of real-world code synthesis~\cite{touvron2023llama,touvron2023llama2}. In this context, leveraging large foundational LLMs such as ChatGPT through \textit{prompting} 
has emerged as a promising avenue for developers seeking to generate high-quality code \cite{brown2020language,li2024acecoder,murr2023testing,ma2023bridging}. However, recent studies have shown that treating LLMs as black-boxes and relying on standard/manual prompting can result in the generation of incorrect code \cite{austin2021program,jain2022jigsaw,sobania2024comparison,liu2024your,spiess2024quality}.

Improving automated prompting approaches is complementary to alternative approaches of improving the code generation capabilities of LLMs. Prior work in this direction include knowledge 
augmentation~\cite{ahmed2024automatic,jain2022jigsaw} and reasoning elicitation~\cite{ma2023bridging,ahmed2023better}. 
Enhanced by self-consistency~\cite{wang2022self}, the correctness of 
generated code repair can be further improved with CoT~\cite{ahmed2023better}. However, it is not clear if similar improvement can be achieved for synthesizing programs from scratch.   
Increasing the specificity of the code-generating prompts can provide additional help~\cite{murr2023testing}, but determining the specificity level is currently a manual process. 

Test-driven software development \cite{beck2003test} is an effective software engineering paradigm where developers 
write tests before code to write correct code. Inspired by the practical success of this paradigm, this paper asks the following question: \textit{Given a base LLM and test cases for a coding task, can we develop an automated prompting approach to generate correct code by minimizing the number of tried prompts (sample-efficiency)?} There are two key challenges in answering this question. First, the search space of prompts is combinatorial and huge. Second, querying the base LLM with a candidate prompt and evaluating the accuracy of code on test cases is expensive. Therefore, we have a challenging search problem.
This paper answers this question by proposing a prompt search approach based on the framework of Bayesian optimization (BO)~\cite{shahriari2016taking}. The key idea behind BO is to learn a surrogate model from the past evaluations (prompt and code accuracy pairs) and use it to intelligently select a sequence of prompts to achieve our goal (generating code with 100 percent accuracy on the given test case). 

Our proposed {\em BO for Code GENeration (BODE-GEN)} approach performs search in continuous embedding space by using an auxiliary LLM to bridge the gap between discrete prompt space and continuous embedding space. The popular Gaussian process (GP) \cite{GP-Book} based surrogate models work well in low dimensions, but the embedding dimension for auxiliary LLM such as LLama2 is 4096 which poses significant challenges. To address this high-dimensional challenge, we leverage two synergistic ideas, namely, random projections and dimensionality scaled priors to build effective GP based surrogate models which are critical for the success of BO. BODE-GEN addresses the core issue of incorrect code generation using LLMs at its source and provides a potential solution for developers. 
\vspace{0.5ex}
\noindent {\bf Contributions.} The key contributions of this paper include: 
\begin{itemize}
    \item Development of a Bayesian optimization approach (BODE-GEN) to iteratively improve prompts for a given base LLM to solve code synthesis tasks.

    \item Demonstrating the effectiveness of BODE-GEN in generating code that meets functional requirements with higher correctness (fraction of passed test cases). 
\end{itemize}

\begin{figure*}
    \centering
    \includegraphics[width=1\textwidth]{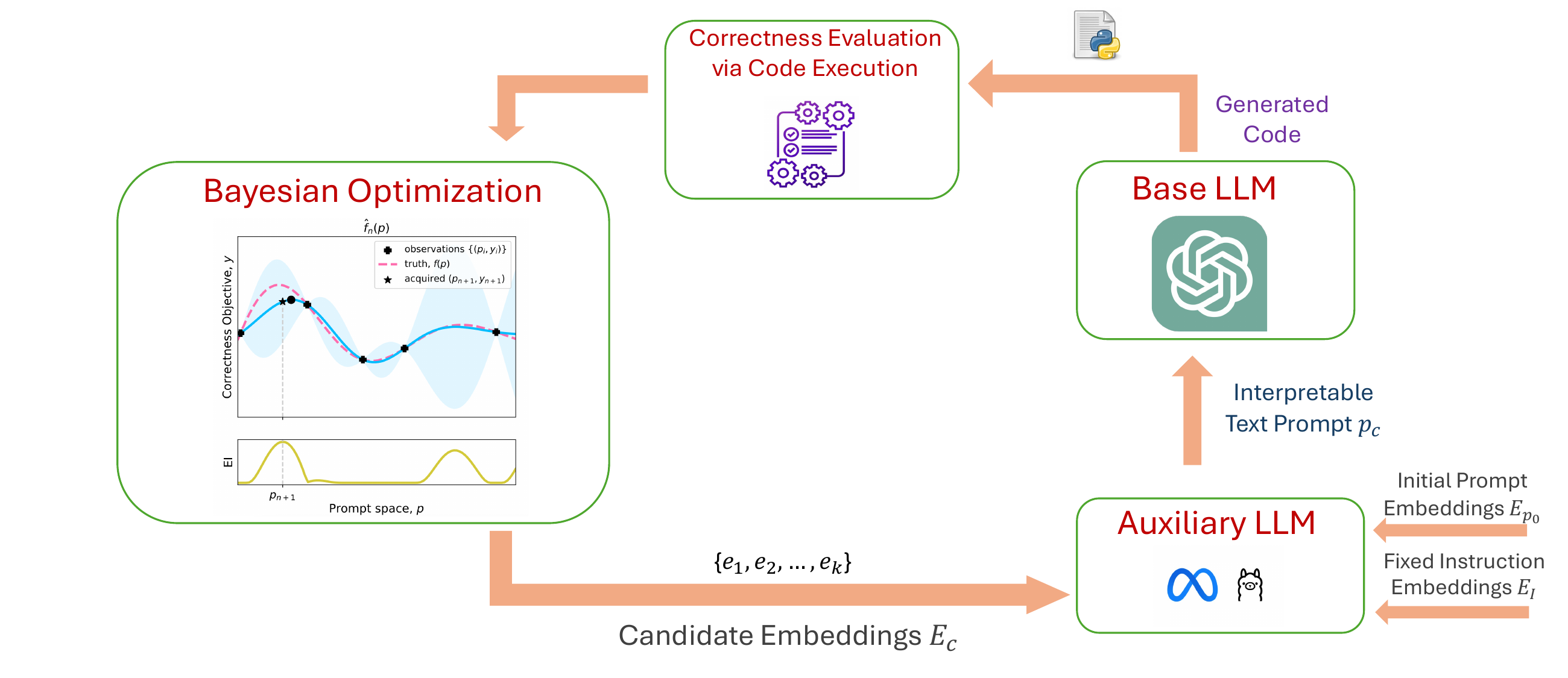}
    \caption{
       High-level overview of our \ourmethod{} approach. The method begins with a set of candidate embeddings $E_c$=$\{e_1, e_2,\cdots, e_m \}$ proposed by the Bayesian Optimization algorithm. These embeddings  combined with the initial prompt embeddings $E_{p_0}$ and fixed instruction embeddings $E_I$ are passed to the auxiliary LLM which generates an interpretable text prompt $p_c$. Subsequently, the base LLM is queried with this prompt $p_c$ to generate the code which is evaluated for functional correctness through code execution on a set of developer provided test cases. The percentage of test cases passed by the code is used as the objective function value for the BO procedure. This overall procedure is repeated for a fixed number of iterations or until we find a prompt that generates code with maximizing code generation accuracy on the given test cases. }
    \label{fig:intro_fig}
\end{figure*}

\section{Problem Setup and Challenges}

Let $LLM_{base}$ denote a base large language model (LLM) that can be queried using textual prompts to solve coding tasks. Given an initial prompt $p_0$ (e.g., \emph{``write Python code for testing whether a given number is prime or not''}) for a coding task $T$ and a set of $n$ developer-provided test cases to verify the correctness of the generated code, our goal is to find a prompt $\hat{p}$ that when used with $LLM_{base}$ will generate functionally correct code, i.e., produces correct outputs on all $n$ test cases.

Suppose Accuracy($LLM_{base}$, $p$) represent the the functional accuracy of the code generated by the given base LLM using the prompt $p$ on $n$ test cases (e.g., 90 percent accuracy means the code passes 90 percent of the given $n$ test cases). Our goal is to find a prompt $\hat{p}$ which maximizes this accuracy and ideally achieves 100 percent accuracy. This problem can be mathematically formulated as follows.
\begin{align}\label{eqn:prompt_opt}
    \hat{p} = \arg \max_p \textsc{Accuracy}(LLM, T)
\end{align}

\vspace{1ex}

\noindent {\bf Key Challenges.} There are two main challenges in solving this optimization problem.

\begin{itemize}

\item {\em Large combinatorial space of prompts.} Each prompt is a sequence of valid tokens. Given a token vocabulary and maximum size of the sequence, the search space of all candidate prompts is combinatorial and very large.

\item {\em Expensive-to-evaluate objective function.} To evaluate the accuracy of a candidate prompt $p$, we need to query the base LLM to generate code and execute it on all $n$ test cases. Each query to base LLM is expensive in terms of dollar cost or computational cost.

\end{itemize}

Therefore, our goal is to minimize the number of queries to the objective function (i.e., the number of tried prompts) to solve this optimization problem. Random search and trial-and-error approaches are not compatible with this goal because their exploration strategy doesn't incorporate machine learning and decision-theoretic reasoning to achieve the target goal.

\section{Bayesian Optimization for Prompt Search}

In this section, we describe an approach for sample-efficient prompt search based on the framework of Bayesian optimization (BO). First, we provide the necessary background on BO. Next, we describe the key challenges in using BO for prompt search and our proposed BODE-GEN approach based on continuous embeddings to address those challenges.

\subsection{Background on Bayesian Optimization}

BO~\cite{shahriari2016taking} is a derivative-free method to \emph{adaptively} and \emph{efficiently} search a given input space $X$ (e.g., search space of prompts) to optimize expensive-to-evaluate objective function $f(x \in X)$. BO is an adaptive procedure because it intelligently selects inputs from $X$ by iterating between querying the objective function $y$=$f(x \in X)$ and making a decision about which input to query next $x_{next}$. BO is sample-efficient because it makes a data-driven decision to select the next input to query the objective function by taking into account all input-output pairs from previous query evaluations. 

Each decision to select the next input from $\mathcal{X}$ to evaluate with $f$ must trade-off two conflicting goals: {\em 1) Exploitation} suggests to use our current, but uncertain, approximation of the input-output relationship, based on the past query evaluations, to select the most promising input in terms of objective function value; and {\em 2) Exploration} suggests to  select the input that  we are most uncertain about to improve our approximation of the input-output relationship.

The key ingredients of BO for data-driven decision making are: 1) {\em surrogate model} that captures our beliefs, based on past objective function evaluations, about the input-output relationship; and 2) \emph{acquisition function} that scores each input according to the utility of querying the objective function on it next. The acquisition function uses the surrogate model of the true input-output relationship $f(x \in X)$ to decide which input to evaluate next by trading-off exploration and exploitation.

The surrogate model $\hat{f}(x)$ is a probabilistic model of the input-output relationship $f(x)$ trained on all input-output pairs from past objective function evaluations. It reflects our current beliefs about $f(x)$ and serves two purposes in BO. First, to guide exploitation, it allows us to cheaply estimate the objective function value of all unevaluated inputs.
Second, to guide exploration, variance quantifies the uncertainties in the predicted objective function value for the unevaluated inputs. 
This makes us aware of regions in input space we need to explore to improve our surrogate model and reduce the uncertainty in our beliefs about $f(x)$. Gaussian processes (GPs) \cite{GP-Book} are the most commonly used surrogate models in BO owing to their flexibility as function approximators and principled uncertainty quantification.

The acquisition function scores the utility of evaluating the next input with the expensive objective function $f$. Here, ``utility'' is defined in terms of our ultimate goal of finding the optimal input with the minimum number of objective function evaluation queries. 
The acquisition function employs the prediction of the objective function value and the associated uncertainty from the surrogate model to assign a utility score to each candidate input that balances exploitation and exploration, respectively. The decision of which input to evaluate next is made by maximizing the acquisition function. Importantly, the acquisition function is cheap to evaluate. Some popular acquisition functions include expected improvement (EI) and upper confidence bound (UCB) \cite{shahriari2016taking}.

To summarize, BO is an iterative procedure that is executed until we reach our goal or maximum iterations are reached. In each iteration, we select the input that maximizes the acquisition function for objective function evaluation and then update the surrogate model based on new training example.

\subsection{BO-based Prompt Search via Continuous Embeddings}

Much of the BO success is on continuous spaces with small number of input dimensions. There are two intertwined surrogate modeling challenges in applying BO for prompt search. First, as opposed to continuous inputs, modeling of combinatorial objects (e.g., sequences) is quite challenging because of a lack of general notion of smoothness on such objects. This is especially exacerbated in the {\em small-data regime} where we have access to only a small number of supervised examples from the input space. Second, the search space of prompts is high-dimensional. We provide principled solutions to address these challenges as part of our proposed BODE-GEN approach and explain their details below.

BODE-GEN performs search for optimized prompts in a continuous embedding space as opposed to the discrete prompt space. The key insight is to leverage an auxiliary open-source LLM $LLM_{aux}$ (e.g., LLaMA 2) to bridge the gap between continuous embedding space and discrete prompt space. Specifically, we perform BO (both surrogate modeling and acquisition function optimization) in the continuous embedding space. The continuous search space for our BO method is parameterized as a set of $d$-dimensional embeddings $E$=$\{e_1, e_2,\cdots, e_m \}$, where each $e_i \in \mathbb{R}^d$ is a continuous vector in the $d$-dimensional embedding space of a local auxiliary LLM $LLM_{aux}$. 

We perform the following sequence of steps in each iteration of BODE-GEN (see Algorithm~\ref{BODEGEN} for pseudo-code and Figure~\ref{fig:intro_fig} for illustration) given a surrogate model trained on the continuous embedding space.

\begin{algorithm}[t!]
\caption{\ourmethod{} Algorithm for Prompt Optimization}
\begin{algorithmic}[1]
\Require Coding task $T$ and $n$ test cases; Initial prompt $p_0$; Base LLM $LLM_{base}$; and Auxiliary LLM $LLM_{aux}$ 
\Ensure Optimized prompt $\hat{p}$

\State $E_{p_0} \gets$ Embedding of initial prompt $p_0$ using $LLM_{aux}$
\State $E_I \gets$ Embedding of fixed instruction using $LLM_{aux}$
\State Initialize surrogate model $\mathcal{M}$ with random projections and dimensionality-scaled priors \cite{hvarfner2024vanilla} on a set of randomly initialized points.

\For{iteration $t = 1$ to $T_{max}$}
    \State $E_c \gets \arg\max_{E} \textsc{AcquisitionFunction}(\mathcal{M}, E)$
    \State $E_{comb} \gets E_I \circ E_c \circ E_{p_0}$  \Comment{Concatenated embedding}
    \State $p_c \gets LLM_{aux}(E_{comb})$ \Comment{Generate discrete prompt}
    \State $C \gets LLM_{base}(p_c)$ \Comment{Generate code using prompt $p_c$}
    \State Acc $\gets \textsc{Evaluate}(C, n)$ \Comment{Accuracy on $n$ test cases}
    
    
    \State Update surrogate model $\mathcal{M}$ with $(E_c, Acc)$
\EndFor

\State \Return best found prompt $\hat{p}$ in terms of code accuracy
\end{algorithmic}
\label{BODEGEN}
\end{algorithm}

\begin{enumerate}
    \item Select the candidate input $E_c$ from the continuous embedding space by maximizing the expected improvement (EI) acquisition function.
    \item The selected embedding $E_c$ is added as a suffix to the continuous embedding $E_{p_0}$ of the initial prompt $p_0$. We also prepend the embeddings $E_{I}$ for a simple instruction $I$: ``\texttt{Your task is to rephrase/reformulate the code prompt given below to achieve a higher score on code generation by a large language model. Please provide the rephrased prompt in one block.}'' 
    given to the auxiliary LLM $LLM_{aux}$ inorder to rephrase the original prompt $p_0$. The resulting combined embedding $E_{comb}$=$E_I \circ E_c \circ E_{p_0}$ where $\circ$ stands for concatenation operation.
    
    \item The combined continuous embedding input $E_{comb}$ is passed to the auxiliary LLM $LLM_{aux}$ to generate a human-interpretable discrete prompt $p_c$.
    \item The discrete prompt $p_c$ is passed as input to the base LLM $LLM_{base}$ to generate code $C$. The code $C$ is executed on all $n$ test cases to measure the functional accuracy, namely, \textsc{Accuracy}($LLM_{base}$, $T$).
    \item If the functional accuracy of code $C$ is 100 percent, we return code $C$ as output. Otherwise, the surrogate model is updated using the new training example: input is the continuous embedding $E_c$ and output is code accuracy.
\end{enumerate}

\noindent {\bf Surrogate Modeling over High-Dimensions.} Gaussian Process (GP) \cite{GP-Book} based surrogate models are commonly used in real-world BO applications with small number of input dimensions (typically less than 50). However, the high-dimensional embedding space of auxiliary LLM poses a significant challenge for standard GP models. For example, the embedding dimension for LLama2 is 4096. GP models that are directly fitted on such a high-dimensional continuous space struggle to generalize, especially when the amount of available supervised data is limited, as is considered in this paper. We apply two synergistic techniques to tackle this challenge: random projections followed by dimensionality-scaled priors \cite{hvarfner2024vanilla} for kernel hyper-parameters.

\begin{itemize}
    \item {\bf Random Projections}: We employ random projections \cite{letham2020re} to reduce the dimensionality of our search space. The key intuition behind this approach is that, in high-dimensional spaces, most of the interesting structure in the data lies in a lower-dimensional manifold. Random projections can capture this structure while preserving important properties of the data, such as pairwise distances between points (as formalized by Johnson-Lindenstrauss lemma \cite{larsen2017optimality}). In the context of our prompt optimization setting, random projections allow us to work with a more manageable representation of the embedding space without excessively reducing the information content.  
    Let $x \in \mathbb{R}^d$ be a point in our original high-dimensional embedding space, where $d$ is large (e.g., 4096 for LLaMA 2). We aim to project this point onto a lower-dimensional space $\mathcal{Z} \in \mathbb{R}^k$, where $k \ll d$. The random projection is defined by a matrix $A \in \mathbb{R}^{k \times d}$, where each entry $a_{ij}$ of matrix $A$ is sampled independently from a standard normal distribution:
    \begin{equation}
    a_{ij} \sim \mathcal{N}(0, 1)
    \end{equation}
    The projected point $z \in \mathcal{Z}$ is then obtained by:
    \begin{equation}
    z =  Ax
    \end{equation}    
    After applying random projections, our GP surrogate model is defined on the low-dimensional space $\mathcal{Z}$. 
    
    \item {\bf Dimensionality scaled priors}: While random projections effectively reduce the dimensionality of our search space, the resulting projected space can still have hundreds of dimensions, posing challenges for standard GP models. GP surrogate models are entirely characterized by the choice of a kernel function (covariance function) $k(z, z')$ that measures the similarity between two input points $z$ and $z'$. The choice of kernel function is critical as it encodes our prior beliefs about the function we are trying to model. Many canonical kernels such as RBF (Radial Basis Function) Kernel and Matern Kernel depend on a lengthscale augmented squared Euclidean distance $d(z, z')$ between $z$ and $z'$ i.e.
    \begin{align}
        d(z, z') = \sum_i^k \left( \frac{z_i - z_i'^2}{l_i^2} \right)
    \end{align}
    The lengthscale $\{l_i\}$ is a critical hyper-parameter that captures the smoothness of functions represented by the kernel. In small supervised data settings (as in our problem), careful prior specification for this parameter is critical to achieve good performance on high-dimensional inputs. In order to address this challenge, we consider the recently proposed idea of  scaling the prior on the lengthscale hyper-parameters of the GP kernel with the square root of the input dimensionality of the search space \cite{hvarfner2024vanilla}.
    Specifically, for a $k$-dimensional input space, dimensionality scaled prior for the length-scale parameter $l_i$ for $i \in \{1, 2,\cdots, k\}$ is described as:
    \begin{equation}
    l_i \sim \text{LogNormal}(\mu = \log(\sqrt{k}), \sigma^2 = 1) 
    \end{equation}      
\end{itemize}

Overall, after applying random projections, our GP surrogate model is defined on the reduced space $\mathcal{Z}$ with dimensionality scaled priors on the lengthscale of the kernel $k(z, z')$.

\section{Experiments and Results}

To evaluate the effectiveness of \ourmethod, we investigate the following research questions (RQs):
\begin{itemize}[leftmargin=3pt,labelsep=2pt]
\item \textbf{RQ1}: \textit{How effective is \ourmethod for code generation?}
\item \textbf{RQ2}: \textit{How does \ourmethod compare with \texttt{CoT} and \texttt{OPRO} prompting methods for code generation?}
\item \textbf{RQ3}: \textit{What changes are introduced by \ourmethod to the intial prompt and how do they affect the resulting code?}
\end{itemize}

In what follows, we first describe our experimental setup including benchmark dataset, details of LLMs, configuration of BODE-GEN and baseline methods, and evaluation methodology. Next, we discuss the results to answer the three RQs.

\subsection{Experimental Setup}

\vspace{1ex}

\noindent {\bf Dataset.} We benchmark our proposed \ourmethod{} approach on the HumanEval+ benchmark \cite{human_eval_plus} which is a recent extension of the widely-adopted HumanEval benchmark \cite{chen2021evaluating} for coding tasks. HumanEval consists of 164 python programming tasks, each containing a function signature and an initial prompt which is written as a docstring. The correctness of each task is measured by evaluating it on a set of pre-specified test cases. HumanEval+ extends the number of test cases in HumanEval tasks by 80x making it a challenging program synthesis benchmark. 

\vspace{1ex}
\noindent {\bf Large Language Models.} We employ three different  $LLM_{\text{base}}$ for code generation: one closed-source ChatGPT 3.5 (Turbo) \cite{brown2020language}, and two open source LLMs CodeLlama (7B) \cite{roziere2023code} and DeepSeek-Coder-33b \cite{guo2024deepseek}.
The instruction-tuned version of LLama2 (7B) \cite{touvron2023llama} is used as the Auxiliary LLM $LLM_{aux}$ for all the experiments.

\vspace{1ex}

\noindent {\bf Computing Machine.} All our experiments were run on a machine featuring an Nvidia A40 GPU. The system is built on an $x86\_64$ architecture with a 32-core AMD EPYC 7573X processor. The machine has 251 GB of memory. 

\vspace{1ex}

\noindent {\bf Configuration of \ourmethod{}.} 
The search space for \ourmethod{} is set to four embeddings, i.e., $E$=$\{e_1, e_2,\cdots, e_4 \}$ where each $e_i \in \mathbb{R}^{4096}$. We randomly project each embedding to a 64 dimensional space resulting in a overall search space of 256 = (4 x 64) dimensions. We ran the BO method for  $T_{max}$=$50$ iterations after initializing the surrogate model with 20 randomly picked embedding points (i.e., BO performance curves are shown for 70 iterations). Expected Improvement was picked as the acquisition function due to its practical success without any hyper-parameters and optimized over a discrete set of 10K points. We ran BO for multiple seeds (five) and present results with mean and error bars. 

\vspace{1ex}

\noindent {\bf Baselines.} We compare \ourmethod{} with two baseline methods for code generation: Zero-shot Chain of Thought (CoT) \cite{wei2022chain} and Optimization by PROmpting (OPRO) \cite{yang2024large}. CoT is a popular prompting approach that is shown to illicit reasoning from LLMs for complex tasks. Unlike the original formulation which requires providing multiple intermediate sequence of steps towards solving the problem, we consider the Zero-shot version where we append the key prompt "Let's think step by step" as a suffix to the initial prompt. OPRO is a recent iterative prompt optimization technique that leverages LLMs to iteratively suggest better candidates conditioned on previously found prompts.

\vspace{1ex}

\noindent {\bf Evaluation Metric.} For each coding task $T$, base language model $LLM_{\text{base}}$ is queried with a candidate prompt to generate python code. We employ the \texttt{pass @1} metric defined below \cite{chen2021evaluating} to evaluate the correctness of the generated code:
\begin{align}
    \texttt{pass @1} = \mathbb{E}_{\text{tasks}} \left[1 - \binom{n - c}{1}/\binom{n}{1}\right]
\end{align}
where we generate $n = 3$ code samples per prompt to reduce the variance of the metric  and $c$ is the number of code samples that pass all unit tests for that task.

\begin{figure}[t!]
    \centering
    \begin{minipage}{0.5\textwidth}
        \centering
        \includegraphics[width=\textwidth]{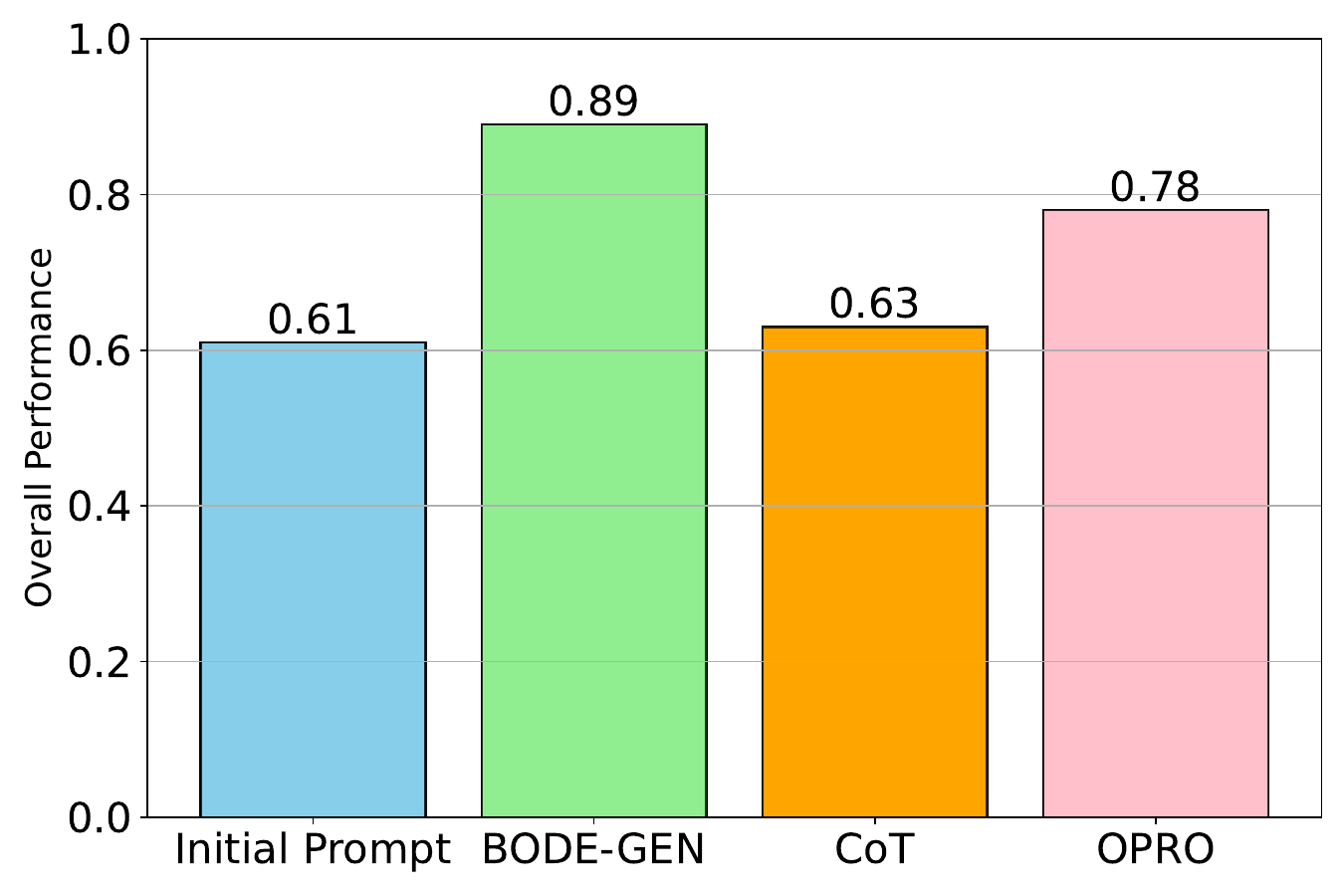}
        \caption*{(a) Results for Base LLM: ChatGPT 3.5(Turbo)}
        \label{fig:op_1}
    \end{minipage}
    \hfill
    \begin{minipage}{0.48\textwidth}
        \centering
        \includegraphics[width=\textwidth]{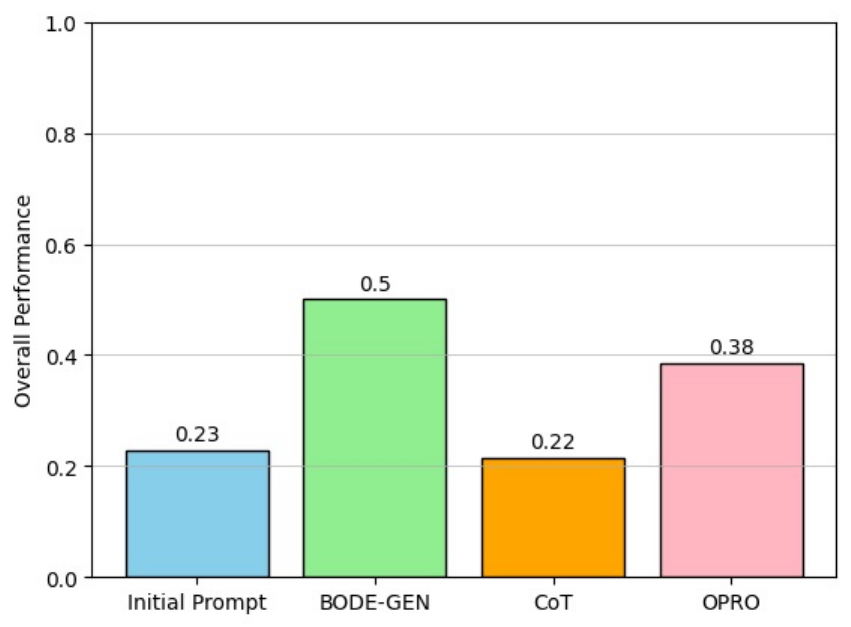}
        \caption*{(b) Results for Base LLM: CodeLlama-7b}
        \label{fig:op_2}
    \end{minipage}
    \caption{Results comparing the overall performance of \ourmethod{} and baselines     with (a) ChatGPT 3.5(Turbo) and (b) CodeLlama-7b as the base LLM. Here, overall performance is computed as the percentage of tasks (out of total 164) from HumanEval+ solved to 100\% correctness.}
    \label{fig:overall_perf_experiments}
\end{figure}

\subsection{RQ1: Effectiveness of \ourmethod{}}
We measure the effectiveness of \ourmethod{} in two ways: the code generation accuracy aggregated over all coding tasks in the HumanEval+ benchmark (higher the better) and the number of BO iterations to achieve high code generation accuracy (smaller the better). \ourmethod{} demonstrates improved performance in finding prompts that improve the generated code's correctness accuracy compared to the baseline methods. The results, as depicted in Figure \ref{fig:overall_perf_experiments}a, show that \ourmethod{} achieves an average code generation accuracy of \textbf{0.89 on ChatGPT-3.5Turbo}\cite{brown2020language}, which is higher than that of initial prompts (0.61), CoT (0.63), and OPRO (0.78). Figure \ref{fig:overall_perf_experiments}b, shows that \ourmethod{} archives an average code generation accuracy of \textbf{0.5 on CodeLlama-7b}\cite{roziere2023code}, which is higher than that of initial prompts (0.23), CoT (0.22), and OPRO (0.38).Figure \ref{fig:DeepseekCoder_baseline_comparison}b, shows that \ourmethod{} archives an average code generation accuracy of \textbf{0.94 on DeepSeker-DeepSeek-Coder-33b}\cite{guo2024deepseek} , which is notably higher than that of initial prompts (0.66), and OPRO (0.78).This increase in performance (code accuracy) suggests the utility of \ourmethod{} in finding prompts that generate functionally correct code on the HumanEval+ benchmark. 

\begin{figure*}[h!]
    \centering
    \includegraphics[width=0.8\linewidth]{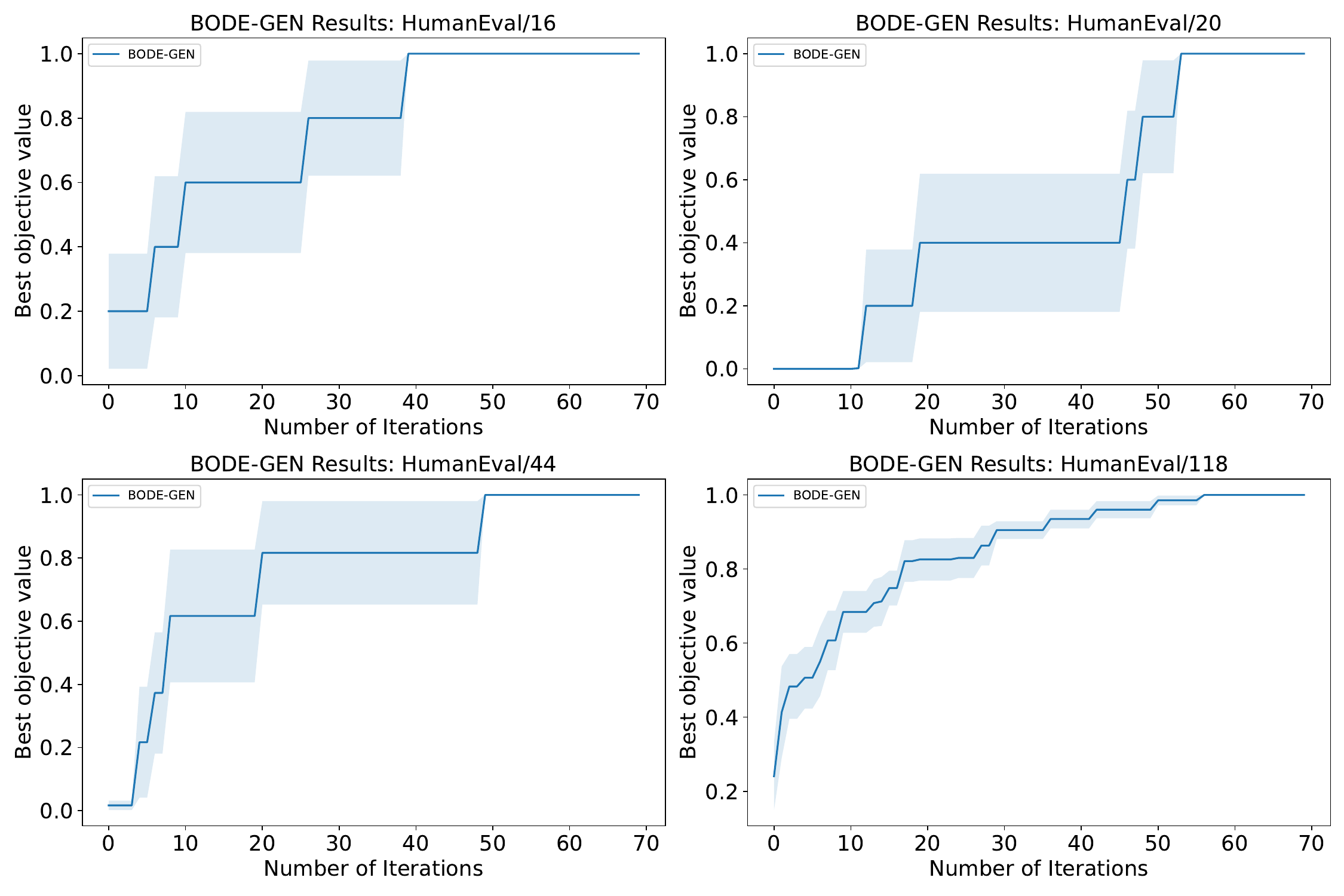}
    \caption{Results showing \ourmethod{}'s performance on ChatGPT 3.5(Turbo) as base-LLM as a function of number of iterations (number of base LLM calls with different prompts) on a subset of representative tasks from HumanEval+ benchmark. 
    Note that the objective value for BO is the percentage of test cases passed by the generated code for a given coding task.  As shown in the figure, prompts suggested by \ourmethod{} are often able to reach high 100\% code generation correctness. Each BO iteration corresponds to roughly one query to the base LLM (precisely it is  three queries per iteration since we generate three samples for each prompt to compute \texttt{pass @1}).}
    \label{fig:bo_curves}
\end{figure*}

\begin{figure*}[t!]
    \centering
    \includegraphics[width=0.8\linewidth]{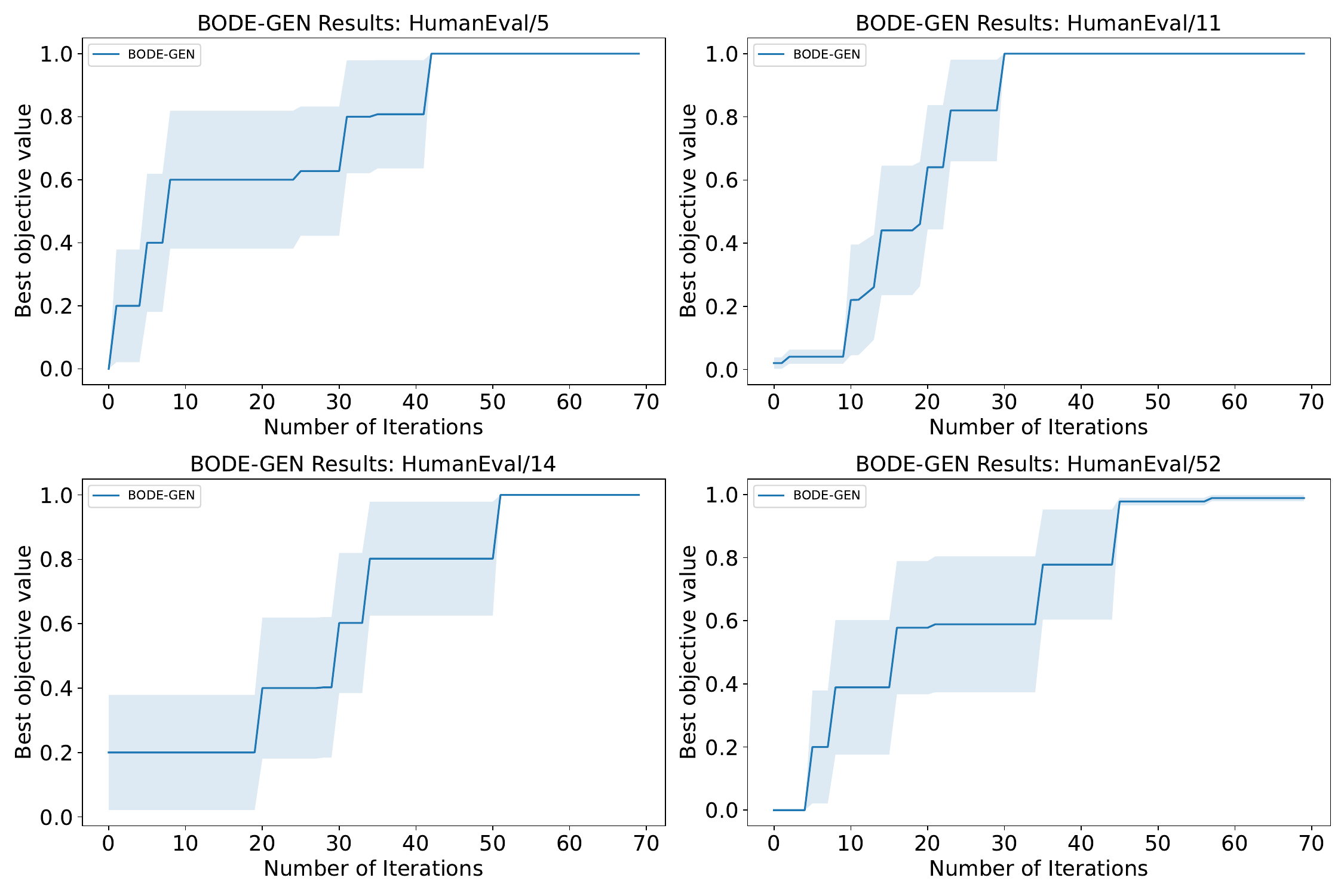}
    \vspace{-10pt}
    \caption{Results showing \ourmethod{}'s performance on CodeLlama-7b as base-LLM as a function of number of iterations (number of base LLM calls with different prompts) on a subset of representative tasks from HumanEval+ benchmark.
    }
    \label{fig:CodeLlama_bo_curves}
\end{figure*}
\begin{figure*}[t!]
    \centering
    \includegraphics[width=0.8\linewidth]{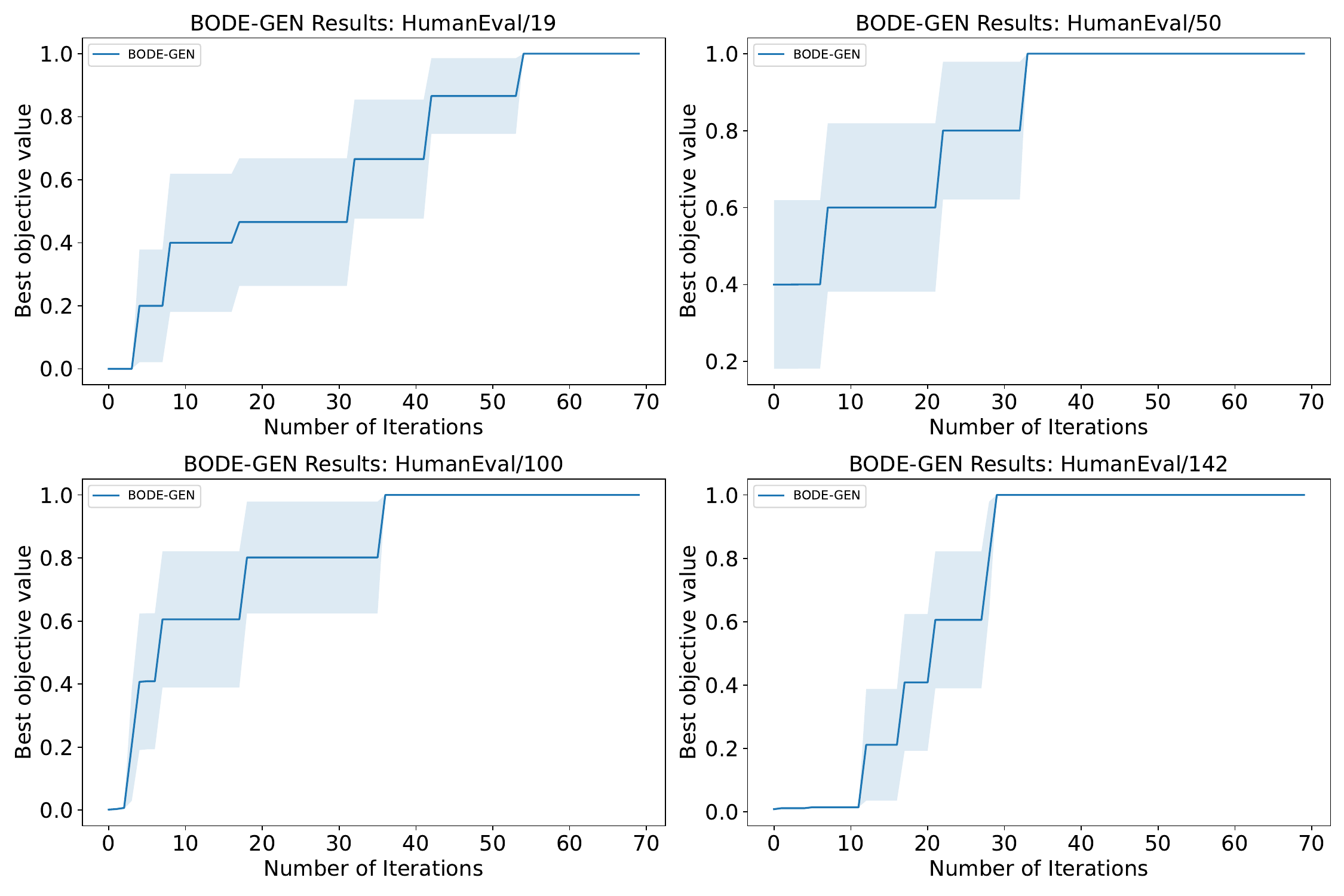}
    \vspace{-10pt}
    \caption{Results showing \ourmethod{}'s performance on DeepSeek-Coder as base-LLM as a function of number of iterations (number of base LLM calls with different prompts) on a subset of representative tasks from HumanEval+ benchmark.}
    \label{fig:DeepseekCoder_bo_curves}
    \vspace{-15pt}
\end{figure*}

We also show the progress of \ourmethod{} in terms of the generated code's accuracy as a function of number of BO iterations (i.e., number of tried prompts) in Figure \ref{fig:bo_curves}, \ref{fig:CodeLlama_bo_curves}, \ref{fig:DeepseekCoder_bo_curves} on some representative coding tasks noting that our findings are similar on other coding tasks. In all cases, \ourmethod{} demonstrates improvement in the objective value (percentage of test cases passed) within the first 20-30 BO iterations, often achieving near-optimal performance by the 50th iteration. This convergence behavior suggests that \ourmethod{} is able to explore the prompt space to find optimized prompts for code generation. As evident from the figures, \ourmethod{} iteratively finds better prompts that are able to reach 100\% correctness accuracy. 

\begin{figure*}[t!]
    
    \centering
    \includegraphics[width=1\linewidth]{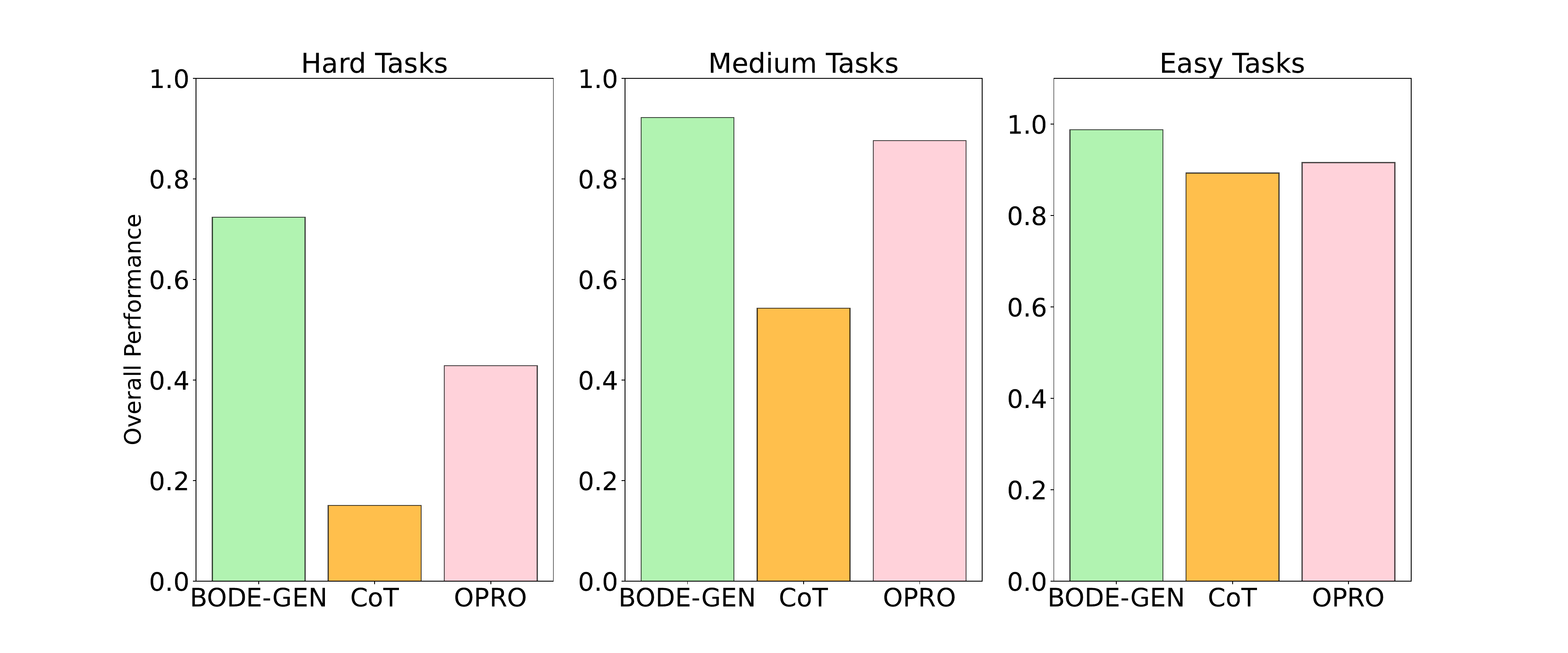}
    \caption{Results comparing the overall performance of \ourmethod{} with zero-shot CoT and OPRO with ChatGPT 3.5(Turbo) as base-LLM on a grouping of HumanEval+ tasks based on a notion of difficulty measured as the correctness of the code generated by initial prompts given for each task. For example, the easy/hard class refers to all tasks for which the code generated via initial prompt achieves correctness (above 67\%/below 30\%) respectively. The medium class contains all tasks with their initial prompts' correctness between 30-67\%.}
    \label{fig:seg_results}
\end{figure*}

\begin{figure}[t!]

    \includegraphics[width=0.8\textwidth]{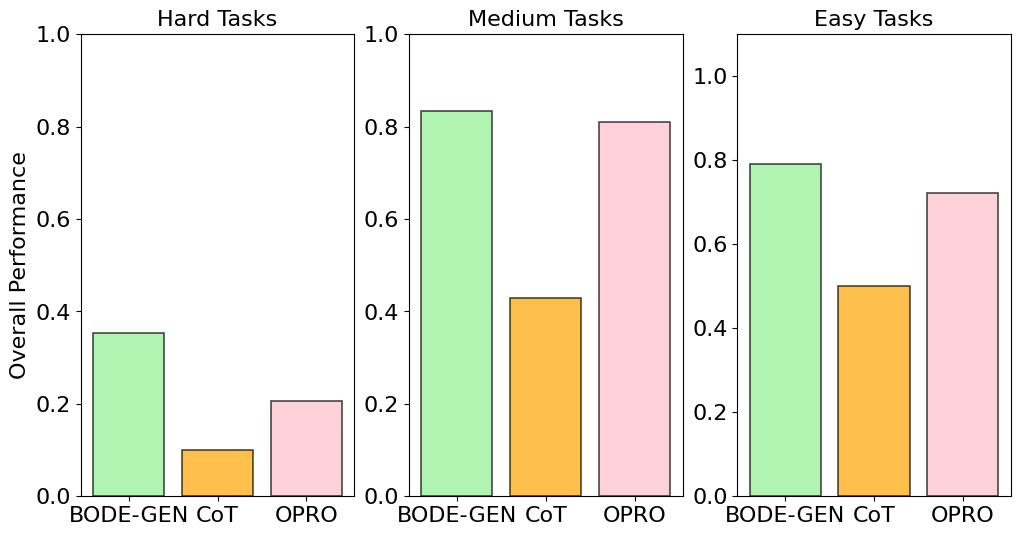}
    \caption{Results of \ourmethod{} and baselines with CodeLlama-7b as base-LLM on a grouping of HumanEval+ tasks based on a notion of difficulty measured as the correctness of the code generated by initial prompts given for each task. For example, the easy/hard class refers to all tasks for which the code generated via initial prompt achieves correctness (above 67\%/below 30\%) respectively. The medium class contains all tasks with their initial prompts' correctness between 30-67\%.}
    \label{fig:dCode_2} 
\end{figure}

\subsection{RQ2: Comparison to Baselines CoT and OPRO}
In comparison to CoT and OPRO, \ourmethod{} outperforms both methods in our setup as shown in Figure \ref{fig:overall_perf_experiments}a, and \ref{fig:overall_perf_experiments}b. CoT shows only small improvement over the initial prompt and performs worse than \ourmethod{}. OPRO achieves better results compared to CoT since it is an iterative approach (similar to \ourmethod{}) that finds better prompts iteratively.
We define a notion of task difficulty as the correctness of the generated code achieved by initial fixed prompt given for each task and create three groups (easy, medium, and hard) of increasing difficult to evaluate all three methods. The results are shown in Figure \ref{fig:seg_results}, \ref{fig:dCode_2} which shows that the gap between \ourmethod{} and the performance of baselines increases as we increase the difficulty of the task. This means that \ourmethod{} is more effective than baselines for hard coding tasks.

\subsection{RQ3: Qualitative Analysis}

To understand how our BO-based prompt optimization approach mproves LLM-driven program synthesis in terms of the correctness of the resulting code, we conducted qualitative analysis on 15 randomly selected coding task cases among those in which the correctness improvement was the most substantial (i.e., the most challenging cases for the LLMs with the \textit{original prompts} used in the HumanEval+ dataset~\cite{liu2024your}). 
By examining these cases, we aim to (1) identify the key changes in the prompt that our optimization approach makes, and (2) based on these changes and how the generated code differs between the initial prompt and optimized prompt, distill common patterns of and main insights into what makes a prompt effective leading LLMs to produce correct code. 

\subsubsection{Key Changes Induced in Prompt Optimizations}
For each of these chosen cases, we carefully compare the two versions of the code-generation prompt, aiming to identify key differences between them in all possible aspects that may affect the base LLM's ability to generate correct code. 

\vspace{0.5ex}

\begin{figure}[h!]
    \centering
    \includegraphics[width=1\linewidth]{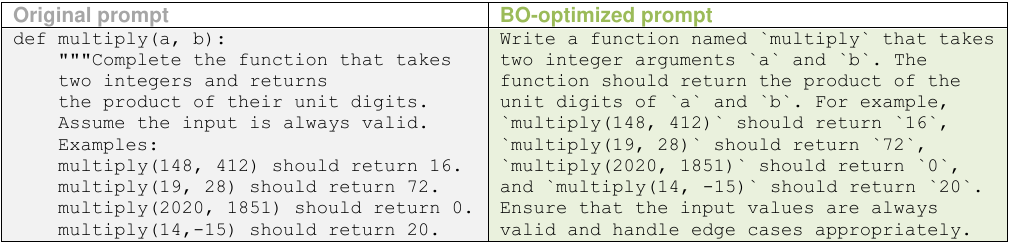}
    \caption{Original prompt versus BO-optimized prompt for generating the {\tt multiply} function (HumanEval+ Case 97).}
    \label{fig:case1}
\end{figure}

\noindent
\textbf{Change 1: Use of Examples.} 
While providing (e.g., input/output) examples generally help LLMs generate correct code, how the examples are used in the prompt matter. 
The original prompts typically include examples within the {\tt docstring}, which might be less visible. 
In optimized prompts from BO, the examples are clearly separated from the main instructions, making them more prominent and easier to reference. 
For example, in the case of the {\tt Multiply} function, shown in Figure~\ref{fig:case1}, the original prompt provides examples within the {\tt docstring}, versus our optimized prompt listing the examples in plain text after describing the task. 
By clearly separating instructions from examples rather than combining them in one information block, these changes improve LLMs' code correctness through \textit{separation of concerns}.

\vspace{0.5ex}

\begin{figure}[h!]
    \centering
    \includegraphics[width=1\linewidth]{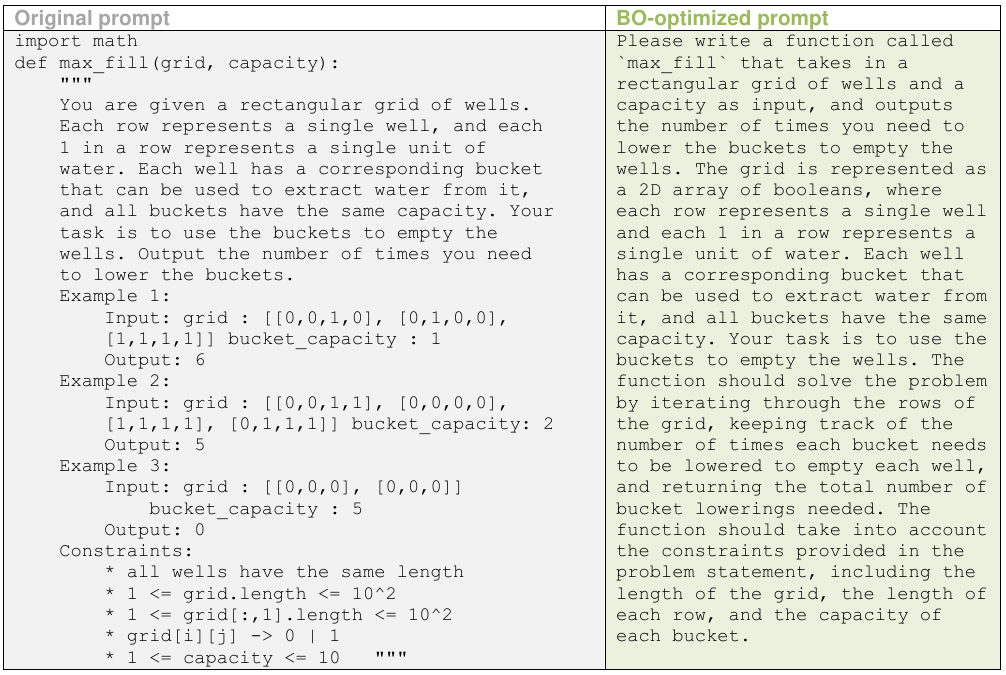}
    \caption{Original prompt versus BO-optimized prompt for generating the {\tt max\_fill} function (HumanEval+ Case 115).}
    \label{fig:case2}
    \vspace{-20pt}
\end{figure}

\noindent
\textbf{Change 2: Instruction Clarity and Detail.}
In several studied cases, the original prompt often includes instructions within the code's {\tt docstring}, which might be concise but less explicit. 
In the optimized version, the prompt provides a detailed, narrative description of the task, often in plain English, making the instructions clearer and more explicit.
For example, in the case of {\tt max\_fill} function generation, as shown in Figure~\ref{fig:case2}, the original prompt (left) describes the task as 
``You are given a rectangular grid of wells...", while the improved prompt provides more details and clarity with ``Please write a function called max\_fill that takes in a rectangular grid of wells and a capacity as input...".

\vspace{0.5ex}

\begin{figure}[h!]
    \centering
    \includegraphics[width=1\linewidth]{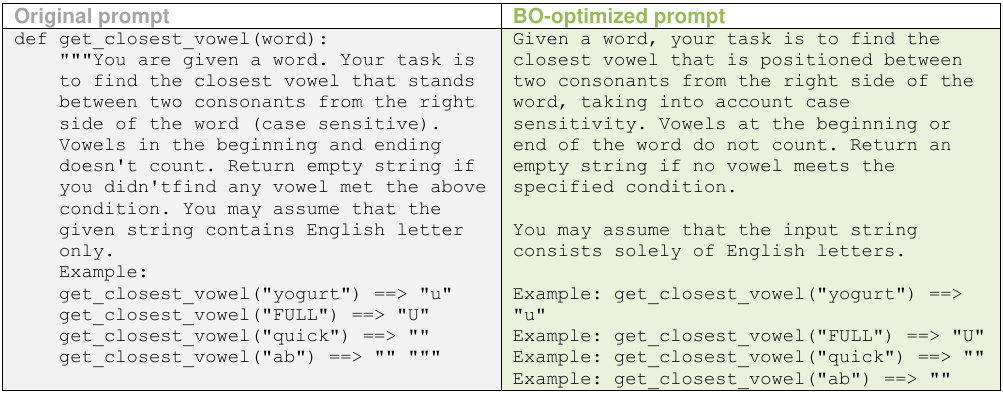}
    \caption{Original prompt versus BO-optimized prompt for generating the {\tt get\_closest\_vowel} function (HumanEval+ Case 118).}
    \label{fig:case3}
\end{figure}

\noindent
\textbf{Change 3: Structure and Organization.}
In most of the studied cases, the original prompts have a less organized structure, while the optimization-resulted prompts use a more structured, step-by-step format, often with clear separations between different parts of the instructions. 
As seen in Figure~\ref{fig:case3}, for generating the {\tt get\_closest\_vowel} function, our optimized prompt improves how the entire prompt is structured: both the assumptions and examples are clearly organized in addition to the task description.

\vspace{0.5ex}

\noindent
\textbf{Change 4: Language and Readability.}
Another main prompt change induced by our optimization lies in 
the language use in the prompt that affects its readability. 
In particular, the original prompts tend to use a technical and formal style typical of in-code documentation, while the optimized prompts use plain English and a more narrative style, improving readability and comprehension. 
Take the {\tt max\_fill} function (Figure~\ref{fig:case2}) as an example again, 
the improved version of the prompt uses a narrative and explanatory style, versus the more technical style of the original prompt. 
Intuitively, LLMs are trained on more natural language corpus than technical documents, which may justify why these changes that improve readability help produce LLMs produce more correct code.

\vspace{0.5ex}

\begin{figure}[h!]
    \centering
    \includegraphics[width=1\linewidth]{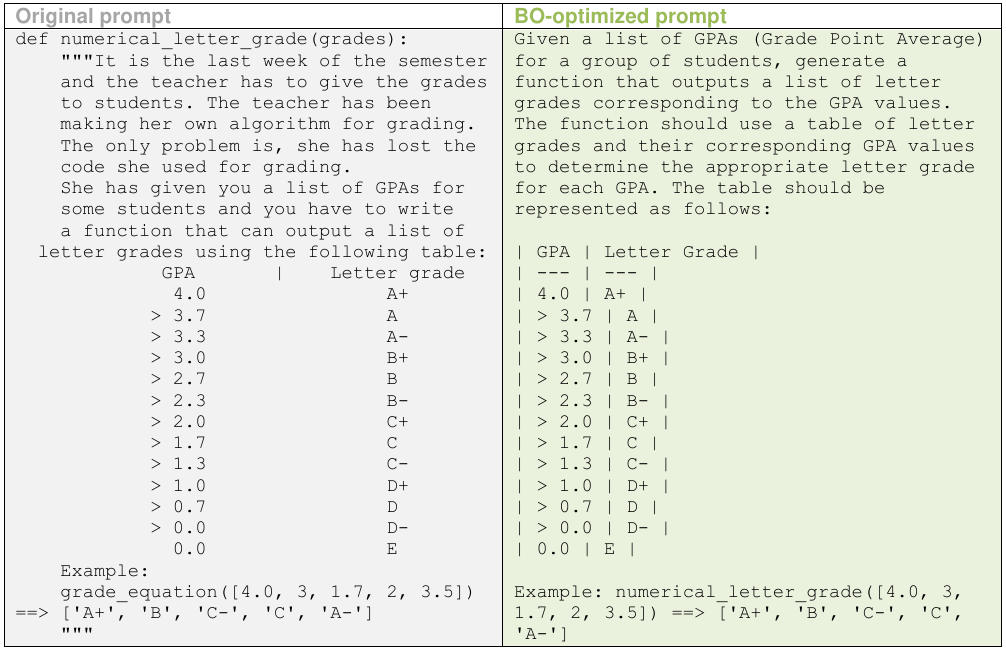}
    \caption{Original prompt versus BO-optimized prompt for generating the {\tt numerical\_letter\_grade} function (HumanEval+ Case 81).}
    \label{fig:case4}
\end{figure}

\noindent
\textbf{Change 5: Descriptive Guidance.}
The original prompts often provide guidance on what the function should do in a general manner. 
In contrast, the optimized prompt has more specific guidance on how to approach the task, including iterative processes or specific logic to follow. 
As shown in Figure~\ref{fig:case2}, if we put aside the detailed examples, the original task description itself is overall general. 
In the optimized version, much more specific guidance is included on iterating through the grid and tracking bucket usage, which helped the LLMs correct the errors in the code generated with the original prompt, even without using those examples. 

To summarize, the common differences between the two versions of each prompt in our studied cases highlight the importance of clarity, detail, explicitness, and structure in writing prompts for code generation. By ensuring that \textit{instructions are clear, detailed, and well-organized}, and by \textit{providing explicit guidance and examples}, the quality and accuracy of the generated code can be significantly improved.

\subsubsection{Patterns of Correct-Code-Generating Prompts}
Based on these prompt changes made by our optimization approach as summarized above, together with comparing the code generated by the two prompt versions, we further identified the following major patterns of correct-code-generating prompts. 

\vspace{0.5ex}

\noindent
\textbf{Pattern 1: Clear and Detailed Instructions.}
We observed that providing detailed and explicit instructions helps LLMs understand the requirements better, leading to more accurate code generation. 
In the case of {\tt max\_fill} function (Figure~\ref{fig:case2}), for example, 
the original prompt states ``Your task is to use the buckets to empty the wells.", which is much less elaborate and explicit than the optimized version: ``The function should solve the problem by iterating through the rows of the grid, keeping track of the number of times each bucket needs to be lowered to empty each well, and returning the total number of bucket lowerings needed."

The \textit{detailed instructions in the optimized prompt clarify the method to solve the problem, resulting in better code generation}. 

\vspace{0.5ex}

\noindent
\textbf{Pattern 2: Step-by-Step Breakdown.}
From multiple cases, it appears clear that breaking down the task into smaller, clear steps helps the model generate code that follows the intended logic more closely. 
One example is found in the prompt for synthesizing the {\tt numerical\_letter\_grade} function, a shown in Figure~\ref{fig:case4}. 
The original version simply provides a context and a list of GPAs, offering no intermediate reasoning. 
The optimized version of the prompt clearly describes the function's purpose, parameters, expected behavior, and gives a table for GPA to letter grade mapping. The \textit{step-by-step breakdown ensures that the model accurately follows the logic needed} to map GPAs to letter grades.

\vspace{0.5ex}

\noindent
\textbf{Pattern 3: Explicit Constraints and Assumptions.}
Inspection of the optimized prompts reveals that specifying constraints and assumptions ensures LLMs adhere to the necessary conditions and edge cases. 
As illustrated in Figure~\ref{fig:case3}, 
the original prompt does provide assumptions and examples but not explicitly (i.e., within a {\tt docstring}), while the optimized version clearly lists the constraints and examples in plain text (i.e., explicitly).
This shows that \textit{explicitly mentioning constraints and assumptions helps the model generate code that respects those conditions}.

\vspace{0.5ex}

\noindent
\textbf{Pattern 4: Separation of Examples and Instructions.}
Almost all of our studied cases confirm that including (e.g., input/output) examples in code-generating prompts help LLMs produce correct code, 
but it is essential to separate examples from instructions. The separation ensures clarity and helps the model focus on understanding both the requirements and the examples. 
As shown in Figure~\ref{fig:case3}, the optimized prompt, which 
lists examples separately after the task description, improves over the original prompt which combines examples and the task description (within the {\tt docstring}). 
Apparently, \textit{clearly separating examples from instructions enhances LLMs' ability to parse and understand both parts effectively}.

\vspace{0.5ex}

\noindent
\textbf{Pattern 5: Plain English Descriptions.}
As evidenced in many cases, using plain English to describe the task makes it easier for LLMs to parse the requirements and generate correct code. 
As seen in Figure~\ref{fig:case2}, the original prompt uses a technical and formal style, while the improved prompt provides a narrative and explanatory style. \textit{Plain English task descriptions make it easier for the model to understand the task and generate correct logic}.

In summary, to write better prompts for code generation with LLMs, it is crucial to \textit{provide clear, detailed instructions, break down tasks into smaller steps, specify constraints and assumptions explicitly, use multiple examples, describe tasks in plain English, and separate examples from instructions}. These patterns help the model understand the requirements better and generate more accurate and correct code.

\begin{figure}
    \centering
    \includegraphics[width=0.5\textwidth]{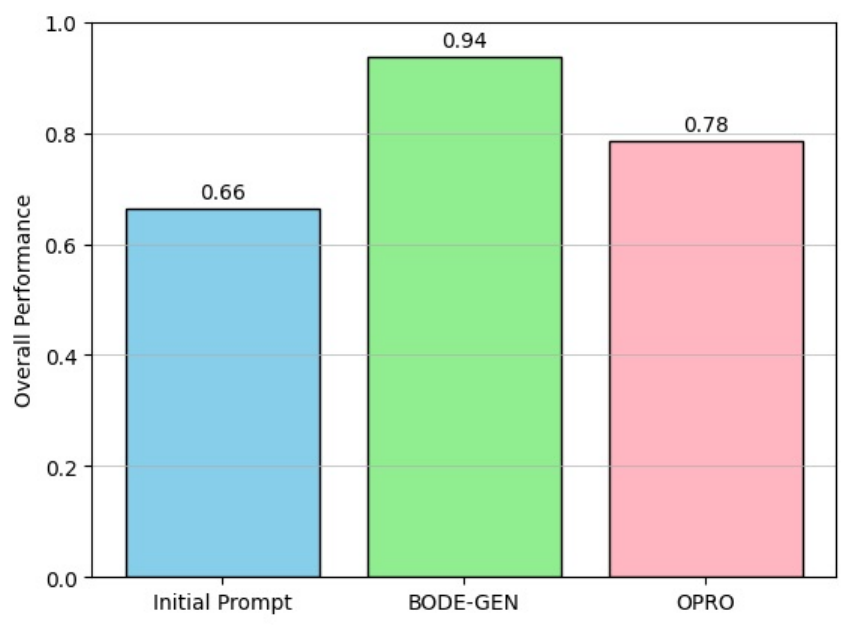}
    \caption{Results comparing the overall performance of \ourmethod{} with other baselines on the HumanEval+ benchmark with DeepSeek-Coder as the base LLM (analogous to Figure \ref{fig:overall_perf_experiments}).}
    \label{fig:DeepseekCoder_baseline_comparison}
\end{figure}

\begin{figure}
    \centering
    \includegraphics[width=0.8\textwidth]{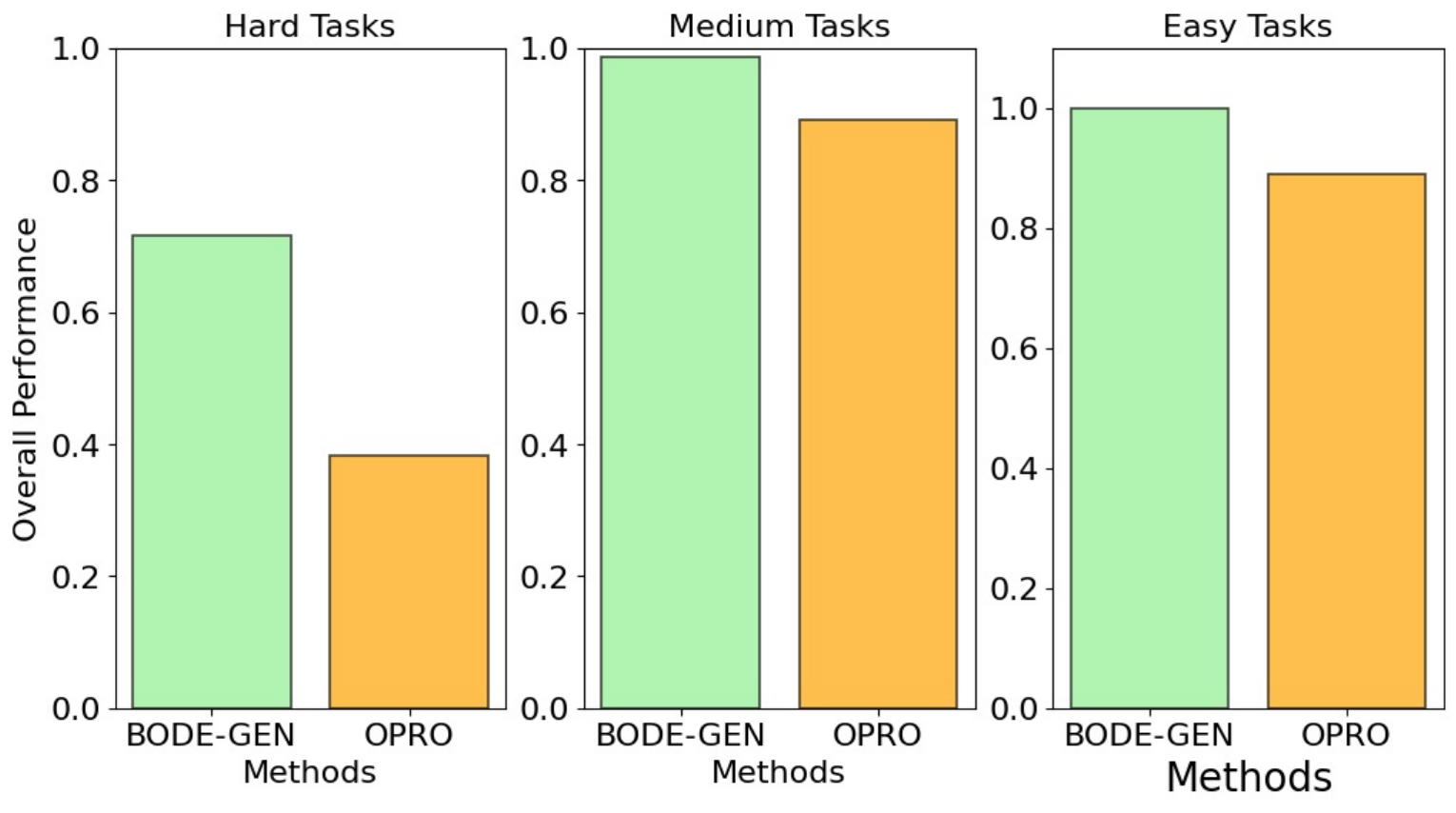}
    \caption{Results comparing the overall performance of \ourmethod{} with OPRO with DeepSeek-Coder as base-LLM on a grouping of HumanEval+ tasks based on a notion of difficulty measured as the correctness of the code generated by initial prompts given for each task (analogous to Figure \ref{fig:overall_perf_experiments}).}
    \label{fig:DeepseekCoder_difficulty_segregation}
\end{figure}

\subsection{Additional Results on DeepSeek-Coder \label{sec:code_llama_results}}
In this section, we present additional results comparing \ourmethod{} with the baseline approaches on DeepSeek-Coder base LLMs. As shown in Figure \ref{fig:DeepseekCoder_baseline_comparison}, with DeepSeek-Coder as base LLM, \ourmethod{} outperforms OPRO in finding prompts that generate code with higher correctness and the gap between the baselines and \ourmethod{} increases on difficult tasks as shown in Figure \ref{fig:DeepseekCoder_difficulty_segregation}.  Since CoT was consistently performing worse than \ourmethod{} and OPRO, we do not include it in this comparison. 
Similar to Figure \ref{fig:bo_curves} in the main paper, we also show the the progress of \ourmethod{} as a function of number of BO iterations with DeepSeek-Coder (Figure \ref{fig:DeepseekCoder_bo_curves}).

\section{Threats to Validity}\label{sec:threats}
The main internal validity threat lies in possible implementation errors in our tool and experimental scripts. To minimize this threat, we have performed careful code reviews of these implementations against manageable testing scenarios. 
Another major issue LLMs are commonly subject to are hallucination~\cite{tian2024codehalu}, which in our study may cause inconsistent and unreliable prompt improvements. To deal with such issues, we set the temperature of the auxiliary LLM to zero in our experiments and compute correctness accuracy  (\texttt{pass @1}) by generating multiple samples from the base LLM. We also ran each experiment multiple times and took results that are consistent among the runs. 
Nevertheless, this mitigation may not have entirely ruled out the possibility that developers may not always get the same correctness improvement when using our technique for prompt optimization, or they may need slightly more iterations to get the same improvement as shown in our evaluation. 

The main external validity threat concerns the datasets we used, as well as LLMs and baseline approaches chosen, in our experiments.  
We used a benchmark (HumanEval+) that is popularly used in LLM evaluations. 
Yet the prompts in it may not represent real-world code-synthesis prompts developers use. 
Meanwhile, we chose capable LLMs that are affordable to use with respect to the scale of our experiments. With the rapid evolution of LLMs, more advanced models may not guaranteed to be improved as much as what we present. 
Similarly, we selected CoT as one of our baselines, which is a state-of-the-art prompting strategy on LLMs. Even more advanced prompting could have achieved better performance (although how to instantiate them for code generation remains an open problem as of now).

\section{Related Work}
\vspace{2pt}\noindent

We provided an overview of the broader literature on LLMs for code generation in the introduction section. Below we discuss closely related work to our specific problem setting.

\vspace{1ex}

\noindent {\bf LLM-driven Code Generation.} To address the challenge that LLMs often generate incorrect code~\cite{austin2021program,liu2024your}, prior work has explored various strategies. One category includes knowledge augmentation~\cite{ahmed2024automatic,jain2022jigsaw} and reasoning elicitation~\cite{ma2023bridging,ahmed2023better}. Chain-of-thoughts (CoT) prompting~\cite{wei2022chain} has been shown to be useful to improve LLM-driven code generation~\cite{ma2023bridging}. However, our experiments on HumanEval benchmark show that the improvement with CoT is small. Jigsaw~\cite{jain2022jigsaw} performs post-processing of the generated code to check and calibrate correctness, by augmenting the LLMs with knowledge about the syntax and semantics of programs based on program analysis and synthesis techniques. This approach is not automatic and maybe incomplete for some coding tasks. 
Ahmed et al.,~\cite{ahmed2023better} leverage the self-consistency technique, which has previously been shown promising for improving the reasoning capabilities of LLMs via CoT, for generating bug fixes by leveraging associated commit logs as explanations. However, it is not clear how to apply this method for synthesizing programs from scratch. Recent work by Murr et al.,~\cite{murr2023testing} found that the specificity of prompts has a major impact on the quality of LLM-generated code: more specific prompts tend to produce accurate code meeting the functional requirements although compromising the generation diversity. However, determining the specificity level is currently a manual process.

A recent prompting technique, AceCoder~\cite{li2024acecoder}, is proposed particularly for code generation with LLMs. It starts with asking the LLM to analyze the given requirements and output an intermediate preliminary output (e.g., test cases), which is then utilized to retrieve similar programs that meet the requirements as exemplars in the code-generation prompt. However, such code examples may not always be available for a given arbitrary requirement.
Following a different strategy, AlphaCodium~\cite{ridnik2024code} is proposed to leverage test-based iterative refinement of the code generation process based on LLMs, where the tests are from what is already available and generated by the AI models.  The effectiveness of AlphaCodium relies on the quality and availability of such tests, which itself is an unresolved challenge. Tao et al.~\cite{tao2024enhancing} employs grammar-guided evolutionary search to improve LLM-based program synthesis. It uses the LLM-generated code as an initial input for the evolutionary search process after a grammar-mapping phase, which allows for program development and fixing errors. Subsequently, it employs different similarity metrics related to the LLM-generated code to steer the multi-objective evolutionary search process. However, the evaluation is only limited to grammar validation of the generated code as opposed to functional correctness. 

Optimization by PROmpting (OPRO) is the state-of-the-art technique for optimizing tasks specified in natural language. OPRO uses LLMs to generate new solutions based on previously evaluated solutions and their scores, iteratively improving the objective function.  Our experiments on HumanEval benchmark demonstrates that OPRO is better than CoT but it is less effective than our proposed BODE-GEN approach in terms of code generation accuracy.

In contrast to prior work, our proposed BODE-GEN approach for optimizing prompts is {\em fully automatic} (no manual intervention) and {\em sample-efficient} (minimized the number of prompts tried) by leveraging the test cases for coding task as per the test-driven software development paradigm \cite{TDD-Book}.

\vspace{1ex}

\noindent {\bf Bayesian Optimization.} BO has shown a lot of success in optimizing continuous spaces with small number of dimensions (typically less than 50). There is relatively less work on BO over discrete spaces which is significantly challenging than BO over continuous spaces \cite{deshwal2021mercer} and is limited to a small number of dimensions. We solve the problem of BO over discrete prompt space by reducing it to BO over high-dimensional continuous embedding space using an auxiliary LLM. We study algorithmic solutions to build effective surrogate models in the small training data setting to handle challenges of the high-dimensional continuous embedding space.

\section{Summary and Future Work}
This paper developed and studied \ourmethod{}, a Bayesian optimization approach for prompt search in large language models (LLMs) for code generation tasks. We formulate this search as optimization over high-dimensional continuous embedding space of an auxiliary LLM. By leveraging random projections and dimensionality-scaled priors, our method addresses the challenges of high dimensionality in the embedding space.  Experiments on the HumanEval+ benchmark demonstrate \ourmethod{}'s effectiveness to significantly improve code generation accuracy across a wide variety of coding tasks when compared to strong baselines. As LLMs continue to play a crucial role in software development, \ourmethod{} offers a framework for improving their reliability and effectiveness, ultimately improving developer's productivity. Future work includes reducing the dependency of \ourmethod{} on the requirement of test cases for coding tasks.



\balance

\bibliographystyle{ACM-Reference-Format}
\bibliography{references}


\begin{thebibliography}{39}


\ifx \showCODEN    \undefined \def \showCODEN     #1{\unskip}     \fi
\ifx \showDOI      \undefined \def \showDOI       #1{#1}\fi
\ifx \showISBNx    \undefined \def \showISBNx     #1{\unskip}     \fi
\ifx \showISBNxiii \undefined \def \showISBNxiii  #1{\unskip}     \fi
\ifx \showISSN     \undefined \def \showISSN      #1{\unskip}     \fi
\ifx \showLCCN     \undefined \def \showLCCN      #1{\unskip}     \fi
\ifx \shownote     \undefined \def \shownote      #1{#1}          \fi
\ifx \showarticletitle \undefined \def \showarticletitle #1{#1}   \fi
\ifx \showURL      \undefined \def \showURL       {\relax}        \fi
\providecommand\bibfield[2]{#2}
\providecommand\bibinfo[2]{#2}
\providecommand\natexlab[1]{#1}
\providecommand\showeprint[2][]{arXiv:#2}

\bibitem[Ahmed and Devanbu(2023)]%
        {ahmed2023better}
\bibfield{author}{\bibinfo{person}{Toufique Ahmed} {and} \bibinfo{person}{Premkumar Devanbu}.} \bibinfo{year}{2023}\natexlab{}.
\newblock \showarticletitle{Better patching using LLM prompting, via Self-Consistency}. In \bibinfo{booktitle}{\emph{2023 38th IEEE/ACM International Conference on Automated Software Engineering (ASE)}}. IEEE, \bibinfo{pages}{1742--1746}.
\newblock


\bibitem[Ahmed et~al\mbox{.}(2024)]%
        {ahmed2024automatic}
\bibfield{author}{\bibinfo{person}{Toufique Ahmed}, \bibinfo{person}{Kunal~Suresh Pai}, \bibinfo{person}{Premkumar Devanbu}, {and} \bibinfo{person}{Earl Barr}.} \bibinfo{year}{2024}\natexlab{}.
\newblock \showarticletitle{Automatic semantic augmentation of language model prompts (for code summarization)}. In \bibinfo{booktitle}{\emph{Proceedings of the IEEE/ACM 46th International Conference on Software Engineering}}. \bibinfo{pages}{1--13}.
\newblock


\bibitem[Austin et~al\mbox{.}(2021)]%
        {austin2021program}
\bibfield{author}{\bibinfo{person}{Jacob Austin}, \bibinfo{person}{Augustus Odena}, \bibinfo{person}{Max Nye}, \bibinfo{person}{Maarten Bosma}, \bibinfo{person}{Henryk Michalewski}, \bibinfo{person}{David Dohan}, \bibinfo{person}{Ellen Jiang}, \bibinfo{person}{Carrie Cai}, \bibinfo{person}{Michael Terry}, \bibinfo{person}{Quoc Le}, {et~al\mbox{.}}} \bibinfo{year}{2021}\natexlab{}.
\newblock \showarticletitle{Program Synthesis with Large Language Models}.
\newblock \bibinfo{journal}{\emph{arXiv preprint arXiv:2108.07732}} (\bibinfo{year}{2021}).
\newblock


\bibitem[Balayn et~al\mbox{.}(2024)]%
        {balayn2024empirical}
\bibfield{author}{\bibinfo{person}{Agathe Balayn}, \bibinfo{person}{Mireia Yurrita}, \bibinfo{person}{Fanny Rancourt}, \bibinfo{person}{Fabio Casati}, {and} \bibinfo{person}{Ujwal Gadiraju}.} \bibinfo{year}{2024}\natexlab{}.
\newblock \showarticletitle{An Empirical Exploration of Trust Dynamics in LLM Supply Chains}.
\newblock \bibinfo{journal}{\emph{arXiv preprint arXiv:2405.16310}} (\bibinfo{year}{2024}).
\newblock


\bibitem[Beck(2002)]%
        {TDD-Book}
\bibfield{author}{\bibinfo{person}{Kent Beck}.} \bibinfo{year}{2002}\natexlab{}.
\newblock \bibinfo{booktitle}{\emph{{Test-Driven Development by Example}}}.
\newblock \bibinfo{publisher}{Addison Wesley. ISBN 978-0-321-14653-3}.
\newblock


\bibitem[Beck(2003)]%
        {beck2003test}
\bibfield{author}{\bibinfo{person}{Kent Beck}.} \bibinfo{year}{2003}\natexlab{}.
\newblock \bibinfo{booktitle}{\emph{Test-driven development: by example}}.
\newblock \bibinfo{publisher}{Addison-Wesley Professional}.
\newblock


\bibitem[Brown et~al\mbox{.}(2020)]%
        {brown2020language}
\bibfield{author}{\bibinfo{person}{Tom~B Brown}, \bibinfo{person}{Benjamin Mann}, \bibinfo{person}{Nick Ryder}, \bibinfo{person}{Melanie Subbiah}, \bibinfo{person}{Jared~D Kaplan}, \bibinfo{person}{Prafulla Dhariwal}, \bibinfo{person}{Arvind Neelakantan}, \bibinfo{person}{Pranav Shyam}, \bibinfo{person}{Girish Sastry}, \bibinfo{person}{Amanda Askell}, {et~al\mbox{.}}} \bibinfo{year}{2020}\natexlab{}.
\newblock \showarticletitle{Language Models are Few-Shot Learners}. In \bibinfo{booktitle}{\emph{Advances in Neural Information Processing Systems}}. \bibinfo{pages}{1877--1901}.
\newblock


\bibitem[Chen et~al\mbox{.}(2021)]%
        {chen2021evaluating}
\bibfield{author}{\bibinfo{person}{Mark Chen}, \bibinfo{person}{Jerry Tworek}, \bibinfo{person}{Heewoo Jun}, \bibinfo{person}{Qiming Yuan}, \bibinfo{person}{Henrique~Ponde de Oliveira~Pinto}, \bibinfo{person}{Jared Kaplan}, \bibinfo{person}{Harri Edwards}, \bibinfo{person}{Yuri Burda}, \bibinfo{person}{Nicholas Joseph}, \bibinfo{person}{Greg Brockman}, \bibinfo{person}{Alex Ray}, \bibinfo{person}{Neel Puri}, \bibinfo{person}{Gretchen Krueger}, \bibinfo{person}{Michael Petrov}, \bibinfo{person}{Heidy Khlaaf}, \bibinfo{person}{Girish Sastry}, \bibinfo{person}{Pamela Mishkin}, \bibinfo{person}{Brooke Chan}, \bibinfo{person}{Scott Gray}, \bibinfo{person}{Nick Ryder}, \bibinfo{person}{Michael Pavlov}, \bibinfo{person}{Alethea Power}, \bibinfo{person}{Lukasz Kaiser}, \bibinfo{person}{Mohammad Bavarian}, \bibinfo{person}{Clemens Winter}, \bibinfo{person}{Phil Tillet}, \bibinfo{person}{Felipe~Petroski Such}, \bibinfo{person}{David Cummings}, \bibinfo{person}{Matthias Plappert}, \bibinfo{person}{Filippos Chantzis},
  \bibinfo{person}{Elizabeth Barnes}, \bibinfo{person}{Ariel Herbert-Voss}, \bibinfo{person}{William~H Guss}, \bibinfo{person}{Alex Nichol}, \bibinfo{person}{Alexander Paino}, \bibinfo{person}{Nikolai Tezak}, \bibinfo{person}{Jie Tang}, \bibinfo{person}{Igor Babuschkin}, \bibinfo{person}{Suchir Balaji}, \bibinfo{person}{Shantanu Jain}, \bibinfo{person}{Andrew Saunders}, \bibinfo{person}{Brandon Houghton}, \bibinfo{person}{Jacob Pfau}, \bibinfo{person}{Diego de Las~Casas}, \bibinfo{person}{Leon Bottou}, \bibinfo{person}{Charles Choi}, \bibinfo{person}{Adam Coates}, \bibinfo{person}{Sam McCandlish}, \bibinfo{person}{Alec Radford}, \bibinfo{person}{Ilya Sutskever}, {and} \bibinfo{person}{Dario Amodei}.} \bibinfo{year}{2021}\natexlab{}.
\newblock \showarticletitle{Evaluating Large Language Models Trained on Code}.
\newblock \bibinfo{journal}{\emph{arXiv preprint arXiv:2107.03374}} (\bibinfo{year}{2021}).
\newblock


\bibitem[Coignion et~al\mbox{.}(2024)]%
        {coignion2024performance}
\bibfield{author}{\bibinfo{person}{Tristan Coignion}, \bibinfo{person}{Cl{\'e}ment Quinton}, {and} \bibinfo{person}{Romain Rouvoy}.} \bibinfo{year}{2024}\natexlab{}.
\newblock \showarticletitle{A Performance Study of LLM-Generated Code on Leetcode}. In \bibinfo{booktitle}{\emph{Proceedings of the 28th International Conference on Evaluation and Assessment in Software Engineering}}. \bibinfo{pages}{79--89}.
\newblock


\bibitem[Deshwal et~al\mbox{.}(2021)]%
        {deshwal2021mercer}
\bibfield{author}{\bibinfo{person}{Aryan Deshwal}, \bibinfo{person}{Syrine Belakaria}, {and} \bibinfo{person}{Janardhan~Rao Doppa}.} \bibinfo{year}{2021}\natexlab{}.
\newblock \showarticletitle{Mercer features for efficient combinatorial Bayesian optimization}. In \bibinfo{booktitle}{\emph{Proceedings of the AAAI Conference on Artificial Intelligence}}, Vol.~\bibinfo{volume}{35}. \bibinfo{pages}{7210--7218}.
\newblock


\bibitem[Ellison et~al\mbox{.}(2010)]%
        {ellison2010evaluating}
\bibfield{author}{\bibinfo{person}{Robert~J Ellison}, \bibinfo{person}{John~B Goodenough}, \bibinfo{person}{Charles~B Weinstock}, {and} \bibinfo{person}{Carol Woody}.} \bibinfo{year}{2010}\natexlab{}.
\newblock \showarticletitle{Evaluating and mitigating software supply chain security risks}.
\newblock \bibinfo{journal}{\emph{Software Engineering Institute, Tech. Rep. CMU/SEI-2010-TN-016}} (\bibinfo{year}{2010}).
\newblock


\bibitem[Enck and Williams(2022)]%
        {enck2022top}
\bibfield{author}{\bibinfo{person}{William Enck} {and} \bibinfo{person}{Laurie Williams}.} \bibinfo{year}{2022}\natexlab{}.
\newblock \showarticletitle{Top five challenges in software supply chain security: Observations from 30 industry and government organizations}.
\newblock \bibinfo{journal}{\emph{IEEE Security \& Privacy}} \bibinfo{volume}{20}, \bibinfo{number}{2} (\bibinfo{year}{2022}), \bibinfo{pages}{96--100}.
\newblock


\bibitem[Gao et~al\mbox{.}(2021)]%
        {gao2021making}
\bibfield{author}{\bibinfo{person}{Tianyu Gao}, \bibinfo{person}{Adam Fisch}, {and} \bibinfo{person}{Danqi Chen}.} \bibinfo{year}{2021}\natexlab{}.
\newblock \showarticletitle{Making Pre-trained Language Models Better Few-shot Learners}. In \bibinfo{booktitle}{\emph{Proceedings of the 59th Annual Meeting of the Association for Computational Linguistics (ACL)}}. \bibinfo{pages}{3816--3830}.
\newblock


\bibitem[Guo et~al\mbox{.}(2024)]%
        {guo2024deepseek}
\bibfield{author}{\bibinfo{person}{Daya Guo}, \bibinfo{person}{Qihao Zhu}, \bibinfo{person}{Dejian Yang}, \bibinfo{person}{Zhenda Xie}, \bibinfo{person}{Kai Dong}, \bibinfo{person}{Wentao Zhang}, \bibinfo{person}{Guanting Chen}, \bibinfo{person}{Xiao Bi}, \bibinfo{person}{Yu Wu}, \bibinfo{person}{YK Li}, {et~al\mbox{.}}} \bibinfo{year}{2024}\natexlab{}.
\newblock \showarticletitle{DeepSeek-Coder: When the Large Language Model Meets Programming--The Rise of Code Intelligence}.
\newblock \bibinfo{journal}{\emph{arXiv preprint arXiv:2401.14196}} (\bibinfo{year}{2024}).
\newblock


\bibitem[Hvarfner et~al\mbox{.}(2024)]%
        {hvarfner2024vanilla}
\bibfield{author}{\bibinfo{person}{Carl Hvarfner}, \bibinfo{person}{Erik~Orm Hellsten}, {and} \bibinfo{person}{Luigi Nardi}.} \bibinfo{year}{2024}\natexlab{}.
\newblock \bibinfo{booktitle}{\emph{Vanilla Bayesian Optimization Performs Great in High Dimensions}}.
\newblock
\urldef\tempurl%
\url{https://openreview.net/forum?id=OfT8MgIqHT}
\showURL{%
\tempurl}


\bibitem[Jain et~al\mbox{.}(2022)]%
        {jain2022jigsaw}
\bibfield{author}{\bibinfo{person}{Naman Jain}, \bibinfo{person}{Skanda Vaidyanath}, \bibinfo{person}{Arun Iyer}, \bibinfo{person}{Nagarajan Natarajan}, \bibinfo{person}{Suresh Parthasarathy}, \bibinfo{person}{Sriram Rajamani}, {and} \bibinfo{person}{Rahul Sharma}.} \bibinfo{year}{2022}\natexlab{}.
\newblock \showarticletitle{Jigsaw: Large language models meet program synthesis}. In \bibinfo{booktitle}{\emph{Proceedings of the 44th International Conference on Software Engineering}}. \bibinfo{pages}{1219--1231}.
\newblock


\bibitem[Larsen and Nelson(2017)]%
        {larsen2017optimality}
\bibfield{author}{\bibinfo{person}{Kasper~Green Larsen} {and} \bibinfo{person}{Jelani Nelson}.} \bibinfo{year}{2017}\natexlab{}.
\newblock \showarticletitle{Optimality of the Johnson-Lindenstrauss lemma}. In \bibinfo{booktitle}{\emph{2017 IEEE 58th Annual Symposium on Foundations of Computer Science (FOCS)}}. IEEE, \bibinfo{pages}{633--638}.
\newblock


\bibitem[Letham et~al\mbox{.}(2020)]%
        {letham2020re}
\bibfield{author}{\bibinfo{person}{Ben Letham}, \bibinfo{person}{Roberto Calandra}, \bibinfo{person}{Akshara Rai}, {and} \bibinfo{person}{Eytan Bakshy}.} \bibinfo{year}{2020}\natexlab{}.
\newblock \showarticletitle{Re-examining linear embeddings for high-dimensional Bayesian optimization}.
\newblock \bibinfo{journal}{\emph{Advances in neural information processing systems}}  \bibinfo{volume}{33} (\bibinfo{year}{2020}), \bibinfo{pages}{1546--1558}.
\newblock


\bibitem[Li et~al\mbox{.}(2024)]%
        {li2024acecoder}
\bibfield{author}{\bibinfo{person}{Jia Li}, \bibinfo{person}{Yunfei Zhao}, \bibinfo{person}{Yongmin Li}, \bibinfo{person}{Ge Li}, {and} \bibinfo{person}{Zhi Jin}.} \bibinfo{year}{2024}\natexlab{}.
\newblock \showarticletitle{AceCoder: An Effective Prompting Technique Specialized in Code Generation}.
\newblock \bibinfo{journal}{\emph{ACM Transactions on Software Engineering and Methodology}} (\bibinfo{year}{2024}).
\newblock


\bibitem[Liu et~al\mbox{.}(2024a)]%
        {liu2024exploring}
\bibfield{author}{\bibinfo{person}{Fang Liu}, \bibinfo{person}{Yang Liu}, \bibinfo{person}{Lin Shi}, \bibinfo{person}{Houkun Huang}, \bibinfo{person}{Ruifeng Wang}, \bibinfo{person}{Zhen Yang}, {and} \bibinfo{person}{Li Zhang}.} \bibinfo{year}{2024}\natexlab{a}.
\newblock \showarticletitle{Exploring and evaluating hallucinations in llm-powered code generation}.
\newblock \bibinfo{journal}{\emph{arXiv preprint arXiv:2404.00971}} (\bibinfo{year}{2024}).
\newblock


\bibitem[Liu et~al\mbox{.}(2024b)]%
        {liu2024your}
\bibfield{author}{\bibinfo{person}{Jiawei Liu}, \bibinfo{person}{Chunqiu~Steven Xia}, \bibinfo{person}{Yuyao Wang}, {and} \bibinfo{person}{Lingming Zhang}.} \bibinfo{year}{2024}\natexlab{b}.
\newblock \showarticletitle{Is your code generated by chatgpt really correct? rigorous evaluation of large language models for code generation}.
\newblock \bibinfo{journal}{\emph{Advances in Neural Information Processing Systems}}  \bibinfo{volume}{36} (\bibinfo{year}{2024}).
\newblock


\bibitem[Liu et~al\mbox{.}(2024c)]%
        {human_eval_plus}
\bibfield{author}{\bibinfo{person}{Jiawei Liu}, \bibinfo{person}{Chunqiu~Steven Xia}, \bibinfo{person}{Yuyao Wang}, {and} \bibinfo{person}{Lingming Zhang}.} \bibinfo{year}{2024}\natexlab{c}.
\newblock \showarticletitle{Is your code generated by chatgpt really correct? rigorous evaluation of large language models for code generation}.
\newblock \bibinfo{journal}{\emph{Advances in Neural Information Processing Systems}}  \bibinfo{volume}{36} (\bibinfo{year}{2024}).
\newblock


\bibitem[Ma et~al\mbox{.}(2023)]%
        {ma2023bridging}
\bibfield{author}{\bibinfo{person}{Yingwei Ma}, \bibinfo{person}{Yue Yu}, \bibinfo{person}{Shanshan Li}, \bibinfo{person}{Yu Jiang}, \bibinfo{person}{Yong Guo}, \bibinfo{person}{Yuanliang Zhang}, \bibinfo{person}{Yutao Xie}, {and} \bibinfo{person}{Xiangke Liao}.} \bibinfo{year}{2023}\natexlab{}.
\newblock \showarticletitle{Bridging Code Semantic and LLMs: Semantic Chain-of-Thought Prompting for Code Generation}.
\newblock \bibinfo{journal}{\emph{arXiv preprint arXiv:2310.10698}} (\bibinfo{year}{2023}).
\newblock


\bibitem[Murr et~al\mbox{.}(2023)]%
        {murr2023testing}
\bibfield{author}{\bibinfo{person}{Lincoln Murr}, \bibinfo{person}{Morgan Grainger}, {and} \bibinfo{person}{David Gao}.} \bibinfo{year}{2023}\natexlab{}.
\newblock \showarticletitle{Testing LLMs on Code Generation with Varying Levels of Prompt Specificity}.
\newblock \bibinfo{journal}{\emph{arXiv preprint arXiv:2311.07599}} (\bibinfo{year}{2023}).
\newblock


\bibitem[Nam et~al\mbox{.}(2024)]%
        {nam2024using}
\bibfield{author}{\bibinfo{person}{Daye Nam}, \bibinfo{person}{Andrew Macvean}, \bibinfo{person}{Vincent Hellendoorn}, \bibinfo{person}{Bogdan Vasilescu}, {and} \bibinfo{person}{Brad Myers}.} \bibinfo{year}{2024}\natexlab{}.
\newblock \showarticletitle{Using an llm to help with code understanding}. In \bibinfo{booktitle}{\emph{Proceedings of the IEEE/ACM 46th International Conference on Software Engineering}}. \bibinfo{pages}{1--13}.
\newblock


\bibitem[Rasmussen and Williams(2006)]%
        {GP-Book}
\bibfield{author}{\bibinfo{person}{Carl~Edward Rasmussen} {and} \bibinfo{person}{Christopher K.~I. Williams}.} \bibinfo{year}{2006}\natexlab{}.
\newblock \bibinfo{booktitle}{\emph{{Gaussian processes for machine learning}}}.
\newblock \bibinfo{publisher}{{MIT} Press}.
\newblock


\bibitem[Ridnik et~al\mbox{.}(2024)]%
        {ridnik2024code}
\bibfield{author}{\bibinfo{person}{Tal Ridnik}, \bibinfo{person}{Dedy Kredo}, {and} \bibinfo{person}{Itamar Friedman}.} \bibinfo{year}{2024}\natexlab{}.
\newblock \showarticletitle{Code generation with alphacodium: From prompt engineering to flow engineering}.
\newblock \bibinfo{journal}{\emph{arXiv preprint arXiv:2401.08500}} (\bibinfo{year}{2024}).
\newblock


\bibitem[Roziere et~al\mbox{.}(2023)]%
        {roziere2023code}
\bibfield{author}{\bibinfo{person}{Baptiste Roziere}, \bibinfo{person}{Jonas Gehring}, \bibinfo{person}{Fabian Gloeckle}, \bibinfo{person}{Sten Sootla}, \bibinfo{person}{Itai Gat}, \bibinfo{person}{Xiaoqing~Ellen Tan}, \bibinfo{person}{Yossi Adi}, \bibinfo{person}{Jingyu Liu}, \bibinfo{person}{Tal Remez}, \bibinfo{person}{J{\'e}r{\'e}my Rapin}, {et~al\mbox{.}}} \bibinfo{year}{2023}\natexlab{}.
\newblock \showarticletitle{Code llama: Open foundation models for code}.
\newblock \bibinfo{journal}{\emph{arXiv preprint arXiv:2308.12950}} (\bibinfo{year}{2023}).
\newblock


\bibitem[Shahriari et~al\mbox{.}(2016)]%
        {shahriari2016taking}
\bibfield{author}{\bibinfo{person}{Bobak Shahriari}, \bibinfo{person}{Kevin Swersky}, \bibinfo{person}{Ziyu Wang}, \bibinfo{person}{Ryan~P Adams}, {and} \bibinfo{person}{Nando De~Freitas}.} \bibinfo{year}{2016}\natexlab{}.
\newblock \showarticletitle{Taking the Human Out of the Loop: A Review of Bayesian Optimization}.
\newblock \bibinfo{journal}{\emph{Proc. IEEE}} \bibinfo{volume}{104}, \bibinfo{number}{1} (\bibinfo{year}{2016}), \bibinfo{pages}{148--175}.
\newblock


\bibitem[Sobania et~al\mbox{.}(2024)]%
        {sobania2024comparison}
\bibfield{author}{\bibinfo{person}{Dominik Sobania}, \bibinfo{person}{Justyna Petke}, \bibinfo{person}{Martin Briesch}, {and} \bibinfo{person}{Franz Rothlauf}.} \bibinfo{year}{2024}\natexlab{}.
\newblock \showarticletitle{A Comparison of Large Language Models and Genetic Programming for Program Synthesis}.
\newblock \bibinfo{journal}{\emph{IEEE Transactions on Evolutionary Computation}} (\bibinfo{year}{2024}).
\newblock


\bibitem[Spiess et~al\mbox{.}(2024)]%
        {spiess2024quality}
\bibfield{author}{\bibinfo{person}{Claudio Spiess}, \bibinfo{person}{David Gros}, \bibinfo{person}{Kunal~Suresh Pai}, \bibinfo{person}{Michael Pradel}, \bibinfo{person}{Md~Rafiqul~Islam Rabin}, \bibinfo{person}{Susmit Jha}, \bibinfo{person}{Prem Devanbu}, {and} \bibinfo{person}{Toufique Ahmed}.} \bibinfo{year}{2024}\natexlab{}.
\newblock \showarticletitle{Quality and Trust in LLM-generated Code}.
\newblock \bibinfo{journal}{\emph{arXiv preprint arXiv:2402.02047}} (\bibinfo{year}{2024}).
\newblock


\bibitem[Tao et~al\mbox{.}(2024)]%
        {tao2024enhancing}
\bibfield{author}{\bibinfo{person}{Ning Tao}, \bibinfo{person}{Anthony Ventresque}, \bibinfo{person}{Vivek Nallur}, {and} \bibinfo{person}{Takfarinas Saber}.} \bibinfo{year}{2024}\natexlab{}.
\newblock \showarticletitle{Enhancing Program Synthesis with Large Language Models Using Many-Objective Grammar-Guided Genetic Programming}.
\newblock \bibinfo{journal}{\emph{Algorithms}} \bibinfo{volume}{17}, \bibinfo{number}{7} (\bibinfo{year}{2024}), \bibinfo{pages}{287}.
\newblock


\bibitem[Tian et~al\mbox{.}(2024)]%
        {tian2024codehalu}
\bibfield{author}{\bibinfo{person}{Yuchen Tian}, \bibinfo{person}{Weixiang Yan}, \bibinfo{person}{Qian Yang}, \bibinfo{person}{Qian Chen}, \bibinfo{person}{Wen Wang}, \bibinfo{person}{Ziyang Luo}, {and} \bibinfo{person}{Lei Ma}.} \bibinfo{year}{2024}\natexlab{}.
\newblock \showarticletitle{CodeHalu: Code Hallucinations in LLMs Driven by Execution-based Verification}.
\newblock \bibinfo{journal}{\emph{arXiv preprint arXiv:2405.00253}} (\bibinfo{year}{2024}).
\newblock


\bibitem[Touvron et~al\mbox{.}(2023a)]%
        {touvron2023llama}
\bibfield{author}{\bibinfo{person}{Hugo Touvron}, \bibinfo{person}{Thibaut Lavril}, \bibinfo{person}{Gautier Izacard}, \bibinfo{person}{Xavier Martinet}, \bibinfo{person}{Marie-Anne Lachaux}, \bibinfo{person}{Timoth{\'e}e Lacroix}, \bibinfo{person}{Baptiste Rozi{\`e}re}, \bibinfo{person}{Naman Goyal}, \bibinfo{person}{Eric Hambro}, \bibinfo{person}{Faisal Azhar}, {et~al\mbox{.}}} \bibinfo{year}{2023}\natexlab{a}.
\newblock \showarticletitle{Llama: Open and efficient foundation language models}.
\newblock \bibinfo{journal}{\emph{arXiv preprint arXiv:2302.13971}} (\bibinfo{year}{2023}).
\newblock


\bibitem[Touvron et~al\mbox{.}(2023b)]%
        {touvron2023llama2}
\bibfield{author}{\bibinfo{person}{Hugo Touvron}, \bibinfo{person}{Louis Martin}, \bibinfo{person}{Kevin Stone}, \bibinfo{person}{Peter Albert}, \bibinfo{person}{Amjad Almahairi}, \bibinfo{person}{Yasmine Babaei}, \bibinfo{person}{Nikolay Bashlykov}, \bibinfo{person}{Soumya Batra}, \bibinfo{person}{Prajjwal Bhargava}, \bibinfo{person}{Shruti Bhosale}, {et~al\mbox{.}}} \bibinfo{year}{2023}\natexlab{b}.
\newblock \showarticletitle{Llama 2: Open foundation and fine-tuned chat models}.
\newblock \bibinfo{journal}{\emph{arXiv preprint arXiv:2307.09288}} (\bibinfo{year}{2023}).
\newblock


\bibitem[Wang et~al\mbox{.}(2021)]%
        {wang2021codex}
\bibfield{author}{\bibinfo{person}{Pengcheng Wang}, \bibinfo{person}{Richard Shin}, \bibinfo{person}{Xiaodong Liu}, \bibinfo{person}{Yao Jin}, \bibinfo{person}{Prafulla Sharma}, \bibinfo{person}{Nitish Keskar}, \bibinfo{person}{Grace Fung}, \bibinfo{person}{Mayur Naik}, {and} \bibinfo{person}{Sameer Yu}.} \bibinfo{year}{2021}\natexlab{}.
\newblock \showarticletitle{Codex: A Large-Scale Neural Network Model for Code Generation}.
\newblock \bibinfo{journal}{\emph{arXiv preprint arXiv:2106.01482}} (\bibinfo{year}{2021}).
\newblock


\bibitem[Wang et~al\mbox{.}(2022)]%
        {wang2022self}
\bibfield{author}{\bibinfo{person}{Xuezhi Wang}, \bibinfo{person}{Jason Wei}, \bibinfo{person}{Dale Schuurmans}, \bibinfo{person}{Quoc Le}, \bibinfo{person}{Ed Chi}, \bibinfo{person}{Sharan Narang}, \bibinfo{person}{Aakanksha Chowdhery}, {and} \bibinfo{person}{Denny Zhou}.} \bibinfo{year}{2022}\natexlab{}.
\newblock \showarticletitle{Self-consistency improves chain of thought reasoning in language models}.
\newblock \bibinfo{journal}{\emph{arXiv preprint arXiv:2203.11171}} (\bibinfo{year}{2022}).
\newblock


\bibitem[Wei et~al\mbox{.}(2022)]%
        {wei2022chain}
\bibfield{author}{\bibinfo{person}{Jason Wei}, \bibinfo{person}{Xuezhi Wang}, \bibinfo{person}{Dale Schuurmans}, \bibinfo{person}{Maarten Bosma}, \bibinfo{person}{Fei Xia}, \bibinfo{person}{Ed Chi}, \bibinfo{person}{Quoc~V Le}, \bibinfo{person}{Denny Zhou}, {et~al\mbox{.}}} \bibinfo{year}{2022}\natexlab{}.
\newblock \showarticletitle{Chain-of-thought prompting elicits reasoning in large language models}.
\newblock \bibinfo{journal}{\emph{Advances in neural information processing systems}}  \bibinfo{volume}{35} (\bibinfo{year}{2022}), \bibinfo{pages}{24824--24837}.
\newblock


\bibitem[Yang et~al\mbox{.}(2024)]%
        {yang2024large}
\bibfield{author}{\bibinfo{person}{Chengrun Yang}, \bibinfo{person}{Xuezhi Wang}, \bibinfo{person}{Yifeng Lu}, \bibinfo{person}{Hanxiao Liu}, \bibinfo{person}{Quoc~V Le}, \bibinfo{person}{Denny Zhou}, {and} \bibinfo{person}{Xinyun Chen}.} \bibinfo{year}{2024}\natexlab{}.
\newblock \showarticletitle{Large Language Models as Optimizers}. In \bibinfo{booktitle}{\emph{The Twelfth International Conference on Learning Representations}}.
\newblock
\urldef\tempurl%
\url{https://openreview.net/forum?id=Bb4VGOWELI}
\showURL{%
\tempurl}


\end{thebibliography}

\end{document}